\documentstyle[aps]{revtex}
\draft
\input epsf
\def\build#1_#2^#3{\mathrel{\mathop{\kern 0pt#1}\limits_{#2}^{#3}}}
\begin{document}
\title{DEEP INELASTIC SCATTERING OF LEPTONS\\ AND HADRONS IN THE
QCD PARTON MODEL\\ AND EXPERIMENTAL TESTS}

%%%%%%%%%%%%%%
\author{C. Bourrely and J. Soffer}

\address{Centre de Physique Th\'eorique
\footnote{Laboratoire propre au CNRS-UPR 7061}, \\ CNRS-Luminy, 
Case 907, \\F-13288 Marseille Cedex 9 - France}
\maketitle
\begin{abstract}
We present the basic aspects of deep inelastic phenomena in the framework
of the QCD parton model. After recalling briefly the standard kinematics,
we discuss the physical interpretation of unpolarized and polarized
structure functions in terms of parton distributions together with several
important sum rules. We also make a rapid survey of the experimental
situation together with phenomenological tests, in particular concerning
various QCD predictions. In the summary, we try to identify some
significant open questions, to clarify where we stand and to see what
we expect to learn in the future.
\end{abstract}

%%%%%%%%%%%%%%%%%%%%%%%%%%%%%%%%%%%%%%%%%%%%%%%%%%%%%%%%%%%%%%%%%%%%%%
\section{Introduction}
\label{intro}
The deep inelastic phenomena
has been extensively studied, for about a quarter of a century,
both theoretically and experimentally. The main purposes of this vast
physics programme were, first to elucidate the internal structure of
the nucleon and later on, to test perturbative Quantum Chromodynamics
(QCD) which generalizes the parton model. At the end of the sixties,
the first measurements on charged lepton deep inelastic scattering (DIS)
at the Stanford Linear Accelerator (SLAC) have shown that the nucleon
is made of hard point-like objects\cite{PAN}, which was the first
evidence for the existence of sub-nuclear particles called
partons\cite{BJO}. Unlike the $ep$ elastic scattering cross section
which drops very rapidly for large momentum transfer, the observed
cross sections for $eN \to e'X$ were much larger than expected, when the
off-shell photon probing the nucleon had a large $Q^2$. Actually these
cross sections were $Q^2$ independent and they obeyed the scaling
behavior predicted by Bjorken\cite{BJO}, whose physical picture was
first given by Feynman\cite{FEY}, in terms of the quark parton model
(QPM). The explanation of scaling was based on the fact that a hard
collision among partons takes place in such a short time, so that partons
behave as if they were free objects. This significant observation led
to the consideration of non-Abelian gauge field theories, which possess
the crucial property of asymptotic freedom, and to propose QCD as the
fundamental theory for strong interactions\cite{POL}. It is now well
established that QCD is an asymptotically free theory and the strong
interaction coupling $\alpha_s(Q^2)$ becomes small when $Q^2$ is large,
{\it i.e.} at short distances. However scale invariance is broken because
of quantum corrections and therefore one expects scaling violations which
are calculable in perturbative QCD.

The outline of this chapter is as follows. In section~\ref{kindis} we discuss
the kinematics of DIS and define the relevant structure functions for
both unpolarized and polarized cases. We provide the physical
interpretation of the structure functions in terms of quark parton
distributions in section~\ref{physinter}. We also give their main features 
from some
recent data on unpolarized and polarized structure functions and we
recall several important sum rules. In section~\ref{audelaqpm} we go 
briefly one step
beyond the QPM to consider scaling violations and to make
phenomenological tests. Finally in section~\ref{scope}, we give a short
summary on significant open questions, and we try to clarify where we stand 
and what we expect to learn in the future.

%%%%%%%%%%%%%%%%%%%%%%%%%%%%%%%%%%%%%%%%%%%%%%%%%%%%%%%%%%%%%%%%%%%%%%%%%%%%%%%
\section{Kinematics of DIS and structure functions}
\label{kindis}

%%%%%%%%%%%%%%%%%%%%%%%%%%%%%%%%%%%%%%%%%%%%%%%%%%%%%%%%%%%%%%%%%%%%%%%%%%%%%%%
\subsection {$ep\to e'X$ unpolarized case}
%\label{unpolep}

\begin{figure}[ht]
%\vskip1in
%\epsfxsize=8.5cm
\epsfysize=8.0cm
\centerline{\epsfbox{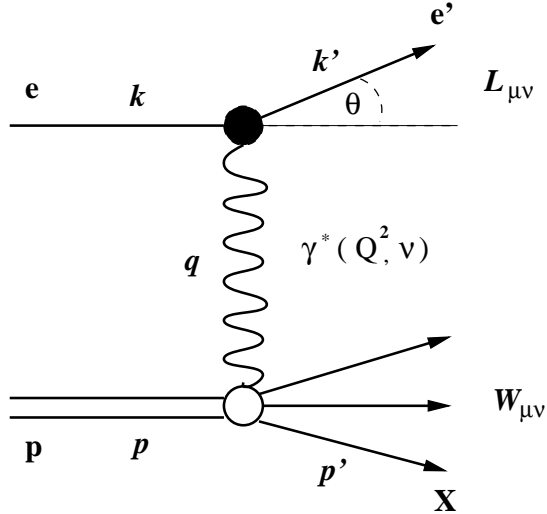}}
\caption{The basic Feynman diagram for deep inelastic electron proton
scattering in lowest order QED. The final hadron state which is not
measured is denoted by $X$.}
\label{fig:graph1}
\end{figure}

Let us start to recall the basic kinematics entering in the
calculation of the cross section for the DIS process $ep\to e'X$ to
lowest order in Quantum Electrodynamics (QED), 
as shown in Fig. \ref{fig:graph1}. 
We define the Lorentz invariants in the usual way
\begin{eqnarray}
s&=&(p+k)^2,\quad q^2=(k-k')^2=-Q^2(Q^2>0),\nonumber \\
\nu&=&pq/M, \quad W^2=p'^2=M^2+2M\nu-Q^2, \label{1}
\end{eqnarray}
where $M$ is the proton mass and we neglect the lepton mass. In the
laboratory frame, if $E$ and $E'$ are the energies of the incoming and
outgoing leptons and $\theta$ is the scattering angle one has
\begin{equation}\label{2}
Q^2=4EE' \sin^2\theta/2\quad\hbox{and}\quad\nu=E-E' ,
\end{equation}
which are the invariant mass and the energy of the exchange off-shell
photon. One also defines two dimensionless scaling variables
\begin{equation}\label{3}
x=Q^2/2p.q=Q^2/2M\nu\quad\hbox{and}\quad y=p.q/p.k=\nu/E ,
\end{equation}
which can vary between zero and one. We recall that $x$ is called the
Bjorken variable and we 
note that $x=1$ corresponds to elastic $ep$ scattering, 
since from eq.(\ref{1}), if $W^2=M^2$ one has $Q^2=2M\nu$.
Moreover for $x=0$, {\it i.e.} $Q^2=0$, we have real photoproduction. Let us
recall that DIS corresponds to the ideal situation where $x$ is fixed
and $Q^2$ and $\nu$ are going to infinity, which is never achieved in practice.
%%%%%%%%%%%%%
%%%%%%%%%
The unpolarized cross section described by the Feynman
diagram shown in Fig. \ref{fig:graph1} has the following expression
\begin{equation}\label{4}
\frac{d^2\sigma}{dE'd\Omega_e}=\frac{4\alpha^2}{Q^4}\frac{E'}{E}
L^{(e)}_{\mu\nu}W^{\mu\nu} ,
\end{equation}
where $\alpha=1/137.036$ is the fine-structure constant, $L^{(e)}_{\mu\nu}$ 
is the symmetric leptonic tensor
\begin{equation}\label{5}
L^{(e)}_{\mu\nu}=2[k'_{\mu}k_{\nu}+k_{\mu}k'_{\nu}+(m^2_e-k.k')g_{\mu\nu}]\ ,
\end{equation}
and where $W_{\mu\nu}$ is the symmetric hadronic tensor defined as
\begin{equation}\label{6}
W_{\mu\nu}=\frac{1}{2}\sum_X \langle p|J^+_{\mu}|X\rangle \langle
X|J_{\nu}|p\rangle (2\pi)^3\delta^{(4)}(p+q-p_X)\ .
\end{equation}
Because the electromagnetic current $J_{\mu}$ is conserved, 
namely $ \partial_{\mu} J_{\mu} = 0$, we have
$q_{\mu}W_{\mu\nu}=q_{\nu}W_{\mu\nu}=0$, and by using parity
conservation and $T$ invariance, one shows that $W_{\mu\nu}$ can be
expressed in terms of two real {\it structure functions} $W_1$ and $W_2$
\begin{eqnarray}
W_{\mu\nu}&=&W_1(\nu,Q^2)\left(-g_{\mu\nu}+\frac{q_{\mu}q_{\nu}}{q^2}
\right) \nonumber \\
&+&\frac{W_2(\nu,Q^2)}{M^2}\left(p_{\mu}-\frac{p.q}{q^2}q_{\mu}\right)
\left(p_{\nu}-\frac{p.q}{q^2}q_{\nu}\right)\ . \label{7}
\end{eqnarray}

%%%%%%%%%%%%%
\begin{figure}[hb]
\epsfxsize=8.0cm
\centerline{\epsfbox{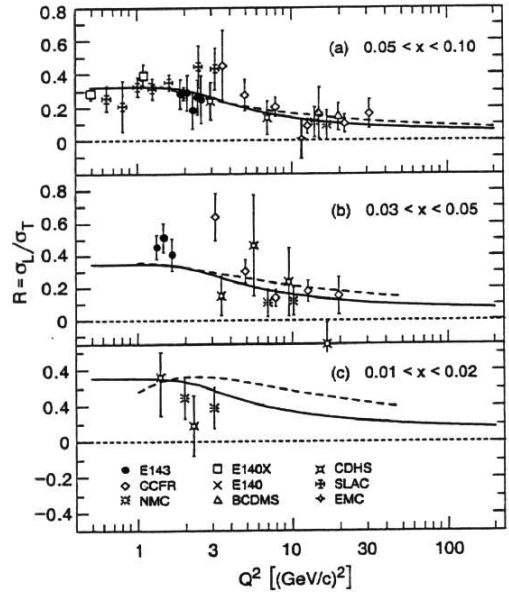}}
\caption{World compilation of R as a function of $Q^2$ for various small $x$
values, curves are a fit of E143 Collaboration\protect\cite{ABE99} . }
\label{fig:rapR}
\end{figure}
%%%%%%%%%
If one neglects the electron mass $m_e$, the cross section eq.(\ref{4})
reads
\begin{equation}\label{8}
\frac{d^2\sigma}{dE'd\Omega_e}=
\frac{4\alpha^2(E')^2}{Q^4}\left[W_2\cos^2\theta/2  +2W_1\sin^2\theta/2
\right]\ ,
\end{equation}
and for the elastic contribution when $Q^2=2M\nu$, {\it i.e.} $x=1$, we
have with $\tau=Q^2/4M^2$,
\begin{eqnarray}
%W^{e\ell}_1&=&\tau G^2_M\delta(\nu-Q^2/2M)\ ,\nonumber \\
W^{e\ell}_1&=&\tau G^2_M\delta(\nu-Q^2/2M)\ , \nonumber\\
W^{e\ell}_2&=&\frac{G^2_E+\tau G^2_M}{1+\tau}\delta(\nu-Q^2/2M)\ , \label{9}
\end{eqnarray}
where $G_E$ and $G_M$ are the electric and magnetic form factors.
Clearly this contribution vanishes in the large $Q^2$ region since
$G_{E,M}(Q^2)\sim(Q^2)^{-2}$ \footnote{This behavior is firmly established
for $G_M(Q^2)$ but recent Jefferson Lab data \cite{JONES} have cast some 
doubts on its validity for $G_E(Q^2)$ .}. 
However for a point-like particle, from eq.(\ref{9}) we have
\begin{equation}\label{10}
2MW_1^{pt}=x\delta(1-x)\quad\hbox{and}\quad \nu W_2^{pt}=\delta(1-x)\ ,
\end{equation}
where $x=Q^2/2M\nu$, and we will come back later to this important
result. From the 1968 SLAC experiment\cite{PAN}, it was observed that at
a fixed value of $W^2$, the invariant square mass of $X$ (see
eq.(\ref{1})), when $Q^2$ is large enough, both $MW_1$ and $\nu W_2$ vanish, 
like in the elastic case. However in the {\it scaling limit}, that is when
both $\nu$ and $Q^2$ are large, with $x=Q^2/2M\nu$ fixed, this is no
longer the case and according to Bjorken \cite{BJO}, one expects
\begin{equation}\label{11}
MW_1(\nu,Q^2)\build{\longrightarrow}_{Q^2\to\infty}^{} F_1(x)\ ,\qquad
\nu W_2(\nu,Q^2)\build{\longrightarrow}_{Q^2\to\infty}^{} F_2(x)\ ,
\end{equation}
where $F_{1,2}(x)$ are two scaling functions independent of $Q^2$.
Therefore the SLAC experiment has shown that the DIS cross section is
much larger than expected and obeys scaling as predicted by Bjorken.
These observations were the basis of the Feynman's idea for the parton
model which can be formulated as follows:
{\it hadrons possess a granular structure
and the ''granules'' behave as hard point-like almost free objects.}

Finally we recall that the structure functions $W_1$ and $W_2$ can be
related to the absorption cross sections for transverse and
longitudinal virtual photons $\sigma_T$ and $\sigma_L$
\begin{eqnarray}
%W_1(\nu,Q^2)=\frac{\sqrt{Q^2+\nu^2}}{4\pi\alpha^2}\sigma_T\ ,\qquad
W_1(\nu,Q^2)&=&\frac{\sqrt{Q^2+\nu^2}}{4\pi\alpha^2}\sigma_T\, \nonumber\\
W_2(\nu,Q^2)&=&\frac{1}{4\pi\alpha^2}\frac{Q^2}{\sqrt{Q^2+\nu^2}}
(\sigma_T+\sigma_L)\ . \label{12}
\end{eqnarray}
This allows to write eq.(8) as
\begin{equation}\label{13}
\frac{d^2\sigma}{dE'd\Omega_e} = \Gamma(\sigma_T+\varepsilon\sigma_L)\ ,
\end{equation}
where $\varepsilon$ and $\Gamma$ are kinematic factors
\begin{eqnarray}
&&\varepsilon=\left[1+2\frac{Q^2+\nu^2}{Q^2}\tan^2\theta/2\right]^{-1}
\quad\hbox{and}\nonumber \\
&&\Gamma=\frac{\sqrt{Q^2+\nu^2}}{2\pi^2Q^2}\frac{E'}{E}
\frac{1}{1-\varepsilon}\ . \label{14}
\end{eqnarray}
The positivity of $\sigma_L$, $\sigma_T$ leads to the inequalities
\begin{equation}\label{15}
0\leq W_1\leq \left(1+\frac{\nu^2}{Q^2}\right)W_2\ .
\end{equation}
One also defines a useful quantity, the ratio $R$ of longitudinal to
transverse cross sections for polarized virtual photons on an unpolarized
target
\begin{equation}\label{16}
R=\sigma_L/\sigma_T=\frac{W_2(\nu,Q^2)}{W_1(\nu,Q^2)}
\left(1+\frac{\nu^2}{Q^2}\right)-1\ .
\end{equation}
 In the limit of large $Q^2$, $\sigma_L$ vanishes, so $R=0$
and we have $2xMW_1=\nu W_2$, that is the so-called Callan-Gross
relation\cite{CAL}
\begin{equation}\label{17}
2xF_1(x)=F_2(x)\ ,
\end{equation}
which is approximately valid for finite $Q^2$. However at finite $Q^2$, 
 quark transverse momentum effects generate non-zero values of $R$. These 
have been observed in the measurements
made at SLAC\cite{WHIT,ABE99} showing that $R$ is slowly varying with $Q^2$ at
fixed $x$. A recent world data compilation is depicted in Fig. \ref{fig:rapR}.

\begin{figure}[ht]
%\vskip1in
\epsfxsize=8.0cm
\epsfysize=10.0cm
\centerline{\epsfbox{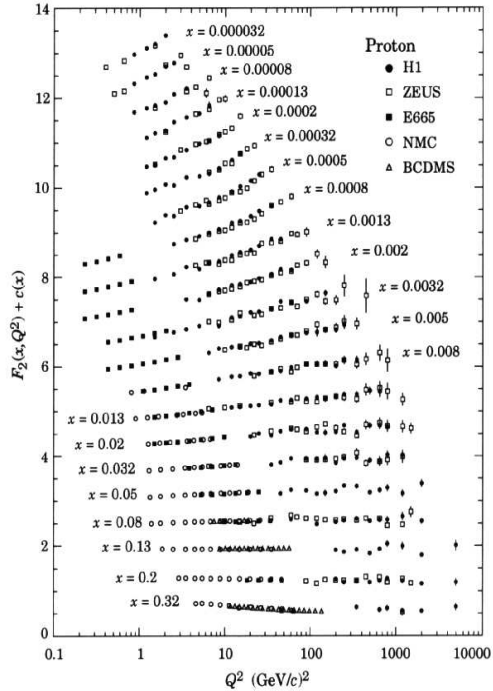}}
\caption{The structure function $F_2^P(x, Q^2)$, data are from 
refs.
\protect\cite{ARN,DER,AHM,BCDMS,E665}. For the purpose of plotting, a constant
$c(x)= 0.6(i_x -0.4)$ is added to $F_{1,2}^p(x,Q^2)$, where $i_x$ is the number
of the $x$ bin ranging from $i_x=1$ ($x=0.32$) to $i_x=21$ ($x=0.000032$) 
(Courtesy of R. Voss).}
\label{fig:f2p}
\end{figure}

Although the two scaling functions $F_{1,2}(x)$ were introduced above in the
limit ${Q^2\to\infty}$, for finite $Q^2$ we have $F_{1,2}(x,Q^2)$ which are
$Q^2$ dependent because of scaling violations. These important aspects will
be discussed in section 0.1.4, but meanwhile in order to complete our
presentation of the data, we give in  Fig. \ref{fig:f2p} a compilation of
$F_{1,2}^p(x,Q^2)$ for fixed proton target experiments (NMC, BCDMS, E665)
and collider experiments at HERA-DESY (ZEUS, H1). Within this broad kinematic 
domain in $x$ and $Q^2$, the data exhibit indeed, a clear evidence 
for scaling violations.
%%%%%%%%%%%%%
%%%%%%%%%

%%%%%%%%%%%%%%%%%%%%%%%%%%%%%%%%%%%%%%%%%%%%%%%%%%%%%%%%%%%%%%%%%%%%%%%%%
\subsection{The neutrino induced reactions $\nu(\bar{\nu})p\to\mu^{\mp} X$}
%\label{nuinduc}

For these reactions the proton is now probed by a weak charged current
and the lowest order diagram is also shown in Fig. \ref{fig:graph1}, where
$\gamma^{\star}$ is replaced by a virtual $W^{\pm}$. Actually,
if $\mu^{\pm}$ are replaced by $e^{\pm}$, one can
also consider the line reversed reactions $e^{\mp}p\to\nu(\bar{\nu})X$
which will be studied at HERA DESY with polarized incident beams $e^{\mp}$.
The cross section is in this case
\begin{equation}\label{18}
\frac{d^2\sigma}{dE'd\Omega}=\frac{G^2_F}{(2\pi)^2}
\left(\frac{M^2_W}{M^2_W+Q^2}\right)^2\frac{E'}{E}L^{(\nu)}_{\mu\nu}
W^{\mu\nu}\ ,
\end{equation}
with $G_F=10^{-5}/M^2$, $M_W=80.43 \mbox{GeV}$ and where the leptonic tensor
$L^{(\nu)}_{\mu\nu}$ is no longer symmetric
\begin{equation}\label{19}
L^{(\nu)}_{\mu\nu}=k'_{\mu}k_{\nu}+k_{\mu}k'_{\nu}-k.k'g_{\mu\nu}\pm
i\varepsilon_{\mu\nu\rho\sigma}k^{\rho}k'^{\sigma}\ ,
\end{equation}
with $+$ for the left-handed neutrino and $-$ for the right-handed
antineutrino. 

\begin{figure}[htbp]
%\vskip1in
\epsfxsize=8.5cm
\epsfysize=12cm
\centerline{\epsfbox{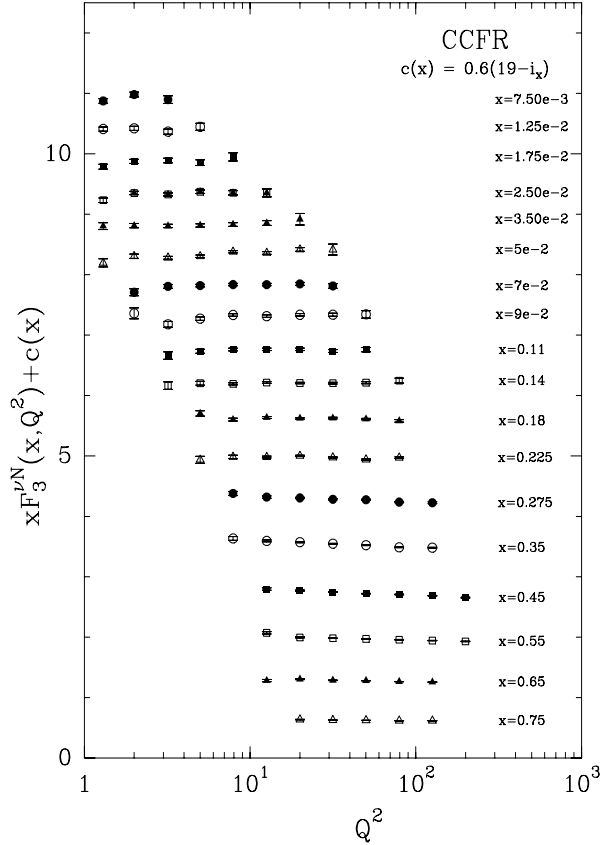}}
\caption{The structure function $x F_3^{\nu N} (x, Q^2)$, where $N$ 
stands for nucleon, versus $Q^2$ for different $x$ values. 
Data are from ref. 
\protect\cite{QUI} and for the purpose of plotting, a constant
$c(x)= 0.6(19 - i_x)$ is added to $x F_3^{\nu N} (x, Q^2)$, where $i_x$ is 
the number of the $x$ bin ranging from $i_x=1$ ($x=0.0075$) to 
$i_x=18$ ($x=0.75$).}
\label{fig:f3ccfr}
\end{figure}

For the hadronic tensor associated to the weak charged current, which is not 
conserved, one has now in addition to the symmetric part
analogous to the previous case,  an antisymmetric term due to the
interference between vector and axial parts of the current.
The hadronic tensor can be expressed in terms of three structure functions
\begin{eqnarray}
W_{\mu\nu}&=&W_1^{\nu}(\nu,Q^2)\left(-g_{\mu\nu}+
{\displaystyle\frac{q_{\mu}q_{\nu}}{q^2}}\right) \nonumber \\
&&
+{\displaystyle\frac{W^{\nu}_2(\nu,Q^2)}{M^2}}\left(p_{\mu}-
{\displaystyle\frac{p.q}{q^2}}q_{\mu}\right)
\left(p_{\nu}-{\displaystyle\frac{p.q}{q^2}q_{\nu}}\right)\cr \nonumber \\
&&
-{\displaystyle\frac{i}{M}}\varepsilon_{\mu\nu\rho\sigma}
p^{\rho}q^{\sigma} W^{\nu}_3(\nu, Q^2)\ . \label{20}
\end{eqnarray}
The new structure function $W^{\nu}_3$, which is not necessarily
positive, is due to the parity violating coupling of the weak gauge
bosons $W^{\pm}$. The cross sections for $\nu$ and $\bar{\nu}$ now read
\begin{eqnarray}
&&\frac{d^2\sigma^{\nu,\bar{\nu}}}{dE'd\Omega}=
\frac{G^2_FE'^2}{2\pi^2(1+Q^2/M^2_W)^2}\cdot \label{21} \\
&&\left[W^{\nu}_2\cos^2\theta/2+2W^{\nu}_1\sin^2\theta/2\mp
\left(\frac{E+E'}{M}\right) W^{\nu}_3\sin^2\theta/2\right] , \nonumber
\end{eqnarray}
and clearly by taking the difference between $\nu$ and $\bar{\nu}$
cross sections one can isolate $W^{\nu}_3$. 
As in the previous case, one can interpret the $W^{\nu}_i(\nu,Q^2)$ 
in terms of the virtual $W$-proton cross sections and from positivity 
one obtains the inequalities
\begin{equation}\label{22}
0\leq\frac{\sqrt{Q^2+\nu^2}}{2M}|W^{\nu}_3|\leq
W^{\nu}_1\leq(1+\nu^2/Q^2)W^{\nu}_2\ .
\end{equation}
Finally one also expects in the scaling limit
\begin{eqnarray}
MW^{\nu}_1(\nu,Q^2)&&\build{\rightarrow}_{Q^2\to\infty}^{}F^{\nu}_1(x), 
\nonumber \\
\nu
W^{\nu}_2(\nu,Q^2)&&\build{\longrightarrow}_{Q^2\to\infty}^{}F^{\nu}_2(x), 
\label{23}\\
\nu
W^{\nu}_3(\nu,Q^2)&&\build{\rightarrow}_{Q^2\to\infty}^{}F^{\nu}_3(x),
\nonumber
\end{eqnarray}
where $F^{\nu}_{1,2,3}(x)$ are three scaling functions independent of
$Q^2$ and similarly to eq.(\ref{17}), one should have
\begin{equation}\label{24}
2xF^{\nu}_1(x)=F^{\nu}_2(x)=x|F^{\nu}_3(x)|\ .
\end{equation}
For finite $Q^2$ these three functions are also $Q^2$ dependent and for
completness we show in Fig. \ref{fig:f3ccfr}, the results on $x F_3^{\nu N}$ 
from the CCFR collaboration
\cite{QUI} at FNAL, as a function of $Q^2$, for fixed $x$ values. 
The existence of scaling violations is also shown by these data.

%%%%%%%%%%%%%
%%%%%%%%%
%%%%%%%%%%%%%%%%%%%%%%%%%%%%%%%%%%%%%%%%%%%%%%%%%%%%%%%%%%%%%%%%%%%%%%%%%%%%%%
\subsection{$ep\to e'X$ polarized case}
%\label{polep}

Let us consider the case of DIS where both electron and proton are
polarized. If $s_e$ denotes the polarization vector of the incident
electron, it is a space-time vector orthogonal to the momentum $k$ of
the electron, so it is such that $k.s_e=0$ and $s^2_e=-1$. Consequently
the leptonic tensor $L_{\mu\nu}$ has now, in addition to the symmetric
part (see eq.(\ref{5})), an antisymmetric part
\begin{equation}\label{25}
L^{(A)}_{\mu\nu}=-2i\varepsilon_{\mu\nu\alpha\beta}
s^{\alpha}_eq^{\beta}\ .
\end{equation}
Similarly, if $s_p$ denotes the polarization vector of the proton the
antisymmetric part of the hadronic tensor is linear in $s_p$ and reads

\begin{eqnarray}
\kern-0,5cm W^{(A)}_{\mu\nu}&=&
i\varepsilon_{\mu\nu\alpha\beta}q^{\alpha}s^{\beta}_pMG_1(\nu,Q^2) \nonumber\\
&&+\frac{i\varepsilon_{\mu\nu\alpha\beta}}{M}
q^{\alpha}\left[(p.q)s^{\beta}_p-(s_p.q)p^{\beta}\right]G_2(\nu,Q^2)\ , 
\label{26}
\end{eqnarray}
where $G_{1,2}$ are two spin-dependent new structure functions. These
structure functions can be related to the absorption cross sections
$\sigma_{1/2}$ and $\sigma_{3/2}$ for virtual photons with projection
$1/2$ and $3/2$ of the total spin along the incident photon direction.
$\sigma_T$ introduced above (see eq.(\ref{12})) is the total transverse
cross section $\sigma_T=\frac{1}{2}(\sigma_{1/2}+\sigma_{3/2})$ and
$\sigma_{TL}$ is the cross section corresponding to the interference
between the transverse and longitudinal polarizations of the photon. 
We have
\begin{eqnarray}
\nu MG_1-Q^2G_2&=&\frac{\sqrt{Q^2+\nu^2}}{2\pi\alpha^2}(\sigma_{1/2}-
\sigma_{3/2})\quad\hbox{and}\nonumber \\
\sqrt{Q^2}(MG_1+\nu G_2)&=&
\frac{\sqrt{Q^2+\nu^2}}{2\pi\alpha^2}\sigma_{TL}\ . \label{27}
\end{eqnarray}
One can also define asymmetries for the virtual photon scattering, namely,
\begin{eqnarray}
A_1 &=&
\frac{\sigma_{1/2}-\sigma_{3/2}}{\sigma_{1/2}+
\sigma_{3/2}}=\frac{\nu MG_1-Q^2G_2}{W_1}\quad\hbox{and} \nonumber\\
A_2 &=&
\frac{2\sigma_{TL}}{\sigma_{1/2}+
\sigma_{3/2}}=\frac{\sqrt{Q^2}( MG_1+\nu G_2)}{W_1}\ , \label{28}
\end{eqnarray}
with the following positivity bounds\cite{DON}
\begin{equation}\label{29}
|A_1|\leq 1\quad\hbox{and}\quad |A_2| \leq\sqrt{R}\ ,
\end{equation}
where $R$ is defined in eq.(\ref{16}). The first inequality is rather trivial
unlike the second one, which can badly restrict the allowed values of $A_2$.
This bound has been recently improved \cite{softer1} as follows
\begin{equation}\label{30a}
|A_2| \leq \sqrt{R(1+A_1)/2}.
\end{equation}
It involves $A_1$ and in the kinematic regions where $A_1$ is small or negative,
one gets a much stronger bound than eq.(\ref{29}) which corresponds to $A_1=1$.
In the scaling limit one also expects
\begin{eqnarray}
\nu M^2G_1(\nu, Q^2)&&\build{\longrightarrow}_{Q^2\to\infty}^{}
g_1(x)\quad \hbox{and} \nonumber\\
\nu^2 MG_2(\nu, Q^2)
&&\build{\longrightarrow}_{Q^2\to\infty}^{} g_2(x)\ ,\label{30}
\end{eqnarray}
where $g_{1,2}(x)$ are two scaling functions independent of $Q^2$. Of
course in this limit, $A_2$ is going to vanish since $R$ also goes to zero.

Finally these asymmetries $A_1$ and $A_2$ can be simply related to the
measured spin asymmetries. In the case of longitudinal beam and target
polarization along the beam direction ({\it i.e.} 
$s_e = \frac{1}{m_e}(|\vec k|, E\vec k/|\vec k|)$ 
and {\it idem} for $s_p$), one has
\begin{equation}\label{31}
A_{\parallel}=\frac{d^2\sigma^{\uparrow\downarrow}-
d^2\sigma^{\uparrow\uparrow}} {d^2\sigma^{\uparrow\downarrow}+
d^2\sigma^{\uparrow\uparrow}}= D(A_1+\eta A_2)\ ,
\end{equation}
where $D$ and $\eta$ are kinematic factors (the factor $D$ is also called
the depolarization factor)
\begin{equation}\label{31a}
D= {1 -E'\varepsilon/E \over 1 + \epsilon R}, \quad
\eta = {\varepsilon\sqrt{Q^2} \over E - E'\epsilon} \ .
\end{equation}
In the region where $\nu$ and $Q^2$ are large, $\eta$ is small and one has
\begin{equation}\label{32}
A_{\parallel}\sim DA_1\ .
\end{equation}
In the case of longitudinally polarized beam and transversely polarized
target ({\it i.e.} $s_p =(0,\vec s)$, $\vec s.\vec p=0$ and $\vec s\ ^2=-1$), 
one has
\begin{equation}\label{33}
A_{\perp}=\frac{d^2\sigma^{\uparrow\rightarrow}-
d^2\sigma^{\uparrow\leftarrow}} {d^2\sigma^{\uparrow\rightarrow}+
d^2\sigma^{\uparrow\leftarrow}} =d(A_2-\xi A_1) ,
\end{equation}
where $d$ and $\xi$ are kinematic factors
\begin{equation}\label{33a}
d = D\sqrt{\frac{2\epsilon}{1+\epsilon}},\quad \xi = \eta {1 + \epsilon \over
2\epsilon} \ .
\end{equation}
When $\nu$ and $Q^2$ are large
\begin{equation}\label{34}
A_{\perp}\sim dA_2\ .
\end{equation}
%%%%%%%%%%%%%
\begin{figure}[htbp]
\epsfxsize=7.5cm
\centerline{\epsfbox{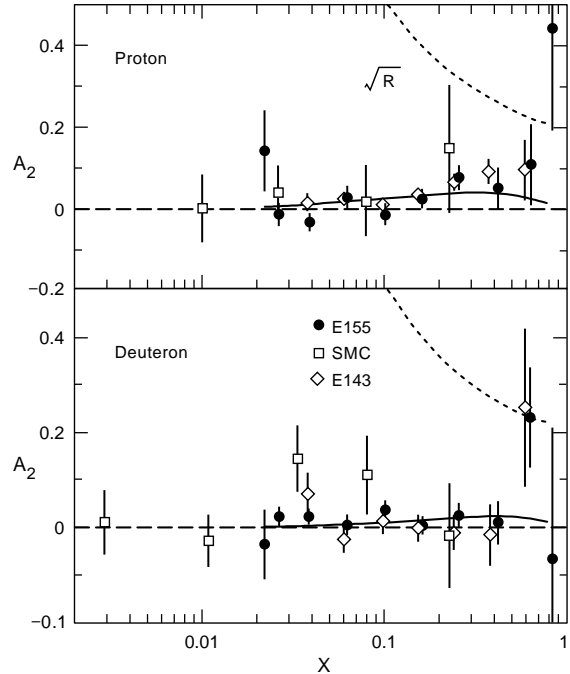}}
\caption{The world data on the asymmetries $A_2$ for proton and deuteron 
targets taken from ref.\protect\cite{Ant99}. The weak positivity bound 
$\sqrt{R}$ is shown as the dotted curve. 
The solid curve was obtained by using $g_2$ 
given by the Wandzura-Wilczek sum rule. (see below eq.(\ref{59}))}
\label{fig:a2e155}
\end{figure}
%%%%%%%%%

\begin{figure}[htbp]
\epsfxsize=8.5cm
\centerline{\epsfbox{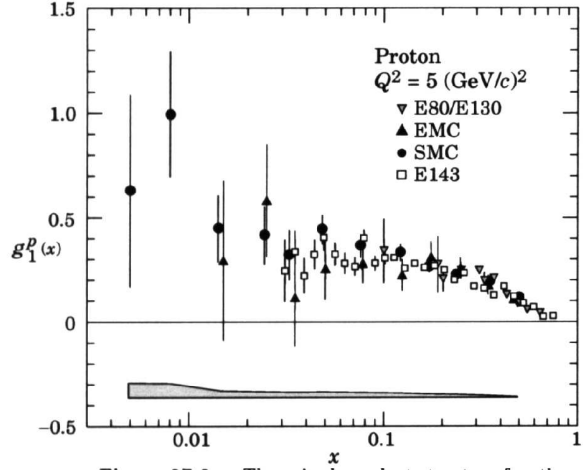}}
\caption{The world data on the structure functions $g_1^p (x)$
from experiments ref.\protect\cite{EMC,SMC,E80,E130,ABE}.}
\label{fig:g1p}
\end{figure}

\begin{figure}[htbp]
\epsfxsize=8.5cm
\centerline{\epsfbox{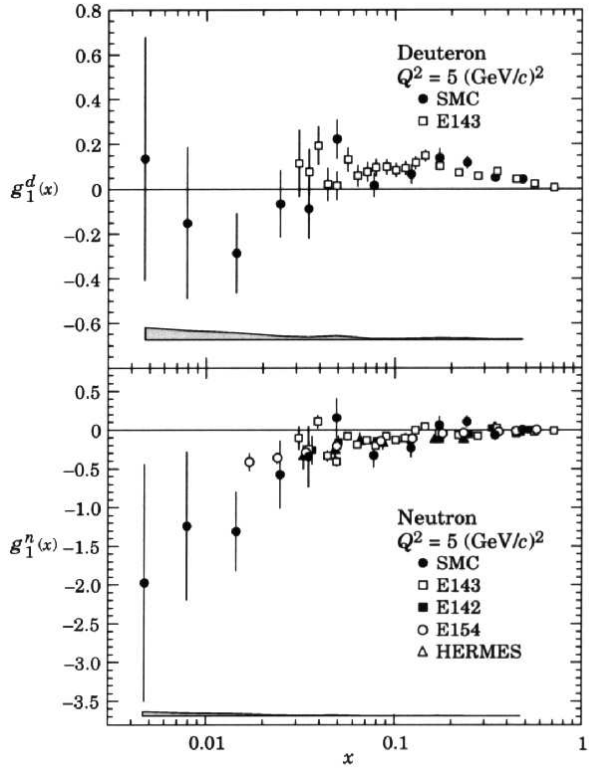}}
\caption{The world data on the structure functions $g_1^d (x)$ and
$g_1^n (x)$ from various experiments ref.
\protect\cite{ABEal,ANT,ADA,E154,E155,HERMES}.}
\label{fig:g1n}
\end{figure}

We show in Fig.\ref{fig:a2e155} a compilation of the world data on
the asymmetries $A_2$ for polarized proton and deuteron targets, which turn out
to be rather small. Experimental results on the asymmetry $A_1$ allow
to extract the function $g_1$, which is also $Q^2$ dependent, 
and we show the most recent data at $Q^2=5GeV^2$, for a polarized proton target
in Fig. \ref{fig:g1p} and for polarized deuterium and neutron targets
\footnote{ Although such a neutron target does not exist, the data are 
obtained from deuterium or from helium 3.} in Fig. \ref{fig:g1n}.

%%%%%%%%%%%%%
%%%%%%%%%
%%%%%%%%%%%%%
%%%%%%%%%
\newpage
%%%%%%%%%%%%%%%%%%%%%%%%%%%%%%%%%%%%%%%%%%%%%%%%%%%%%%%%%%%%%%%%%%%%%%%%%%%%%
\section{Physical interpretation of structure functions, parton
distributions and sum rules}
\label{physinter}

%%%%%%%%%%%%%%%%%%%%%%%%%%%%%%%%%%%%%%%%%%%%%%%%%%%%%%%%%%%%%%%%%%%%%%%%%%%%%
\subsection{The physical picture : the quark parton model}
%\label{physpic}

The experimental facts observed at SLAC in 1968\cite{PAN} gave a strong
support to the existence of point-like objects in the proton structure
and were the inspiration for trying to elaborate a simple physical
picture for interpreting the structure functions of DIS. When $E\to
\infty$ (i.e. $\nu\to\infty$), because of time dilatation, the life
time of the proton virtual states is also very large. When $Q^2\to
\infty$, the interaction time $1/\sqrt{Q^2}$ becomes very small. So in
the scaling limit, $\nu$ and $Q^2$ very large, the virtual photon
$\gamma^{\star}$ (see Fig. \ref{fig:graph1}) sees a ''frozen''
state of quasi-real, quasi-free point like objects called partons.

%%%%%%%%%%%%%
\begin{figure}[htbp]
\epsfxsize=7.0cm
\centerline{\epsfbox{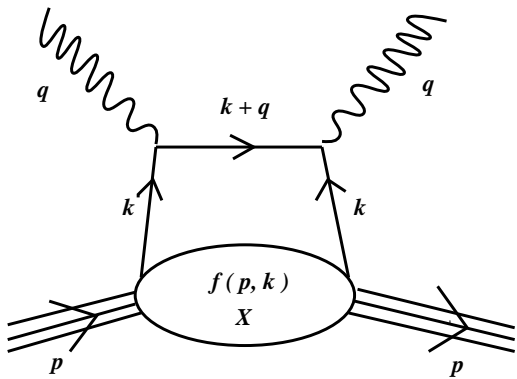}}
\caption{The parton model handbag diagram showing the interaction of
$\gamma^\star$ with the parton of momentum $k$.}
\label{fig:graph2}
\end{figure}
%%%%%%%%%
 
According to the parton model, 
in this kinematical region, the proton behaves as a gas of free
partons. In particular, it is very {\it unlikely} to recover a proton in the
final state and this is why the elastic form factors $G_{E,M}$ vanish
when $Q^2\to\infty$, as recalled above. So in the calculation of the
DIS cross section ({\it e.g.} see eq.(\ref{4})), the hadronic tensor
$W_{\mu\nu}$ is evaluated according to the parton model graph, shown in
Fig. \ref{fig:graph2}, namely
\begin{equation}\label{35}
W_{\mu\nu}=\sum_{i}\sum_{s}\int d^4k f^i_s(p,k)w^i_{\mu\nu}(q,k)\delta
[(k+q)^2] ,
\end{equation}
where $f^i_s(p,k)$ is the probability for the virtual photon to struck
a parton $i$ of spin $s$ and momentum $k$, inside the proton of
momentum $p$. It seems very natural nowdays to assume that the charged
partons which interact directly with $\gamma^{\star}$ are
quarks\footnote{Gluons are also fundamental constituents of the proton,
but since they don't carry electric charge, they don't couple directly
to $\gamma^{\star}$.}. 
This step was not so obvious to make in 1968, although
we had already the $SU(3)$ classification of hadrons, as built up from
quarks, and the Cabibbo theory of weak interactions simply formulated
in terms of leptons and quarks. It is important to recall that nowdays
{\it six} quarks have been identified. They have very different masses:
three of them are the light quarks {\it up}, {\it down} and {\it strange} 
($u,d,s$) and three are the heavy quarks {\it charm}, {\it bottom} and
{\it top} ($c,b,t$). The quark {\it flavor} corresponds to one of these six 
labels $u,d,s,c,b,t$ and the $u,c,t$ quarks have a charge $e=2/3$, while
the charge of the $d,s,b$ quarks is $e=-1/3$. The summation over $i$ in
eq.(\ref{35})) and subsequent equations stands for a flavor summation 
$i=u,d,s,..$.
In the nucleon $u$ and $d$ quarks dominate and their mass can be neglected
in high energy DIS. 
So if these partons are indeed spin
$1/2$ massless quarks, we are now in the framework of the quark parton
model (QPM), $w^i_{\mu\nu}(q,k)$ is the tensor which 
describes the interaction of $\gamma^{\star}$ with the quark $i$ 
of momentum $k$ and we have, for its symmetric part
\begin{equation}\label{36}
w^i_{\mu\nu}(q,k)=\frac{e^2_i}{2} Tr\left\{\rlap/k\gamma_{\mu}(
\rlap/k+\rlap/q)\gamma_{\nu}\right\}\ ,
\end{equation}
which is similar to the leptonic tensor ({\it e.g.} see eq.(\ref{5})). If
$x_i$ is the fraction of the proton momentum carried by the quark $i$,
$x_i=k.q/M\nu$, so
$\delta[(k+q)^2]=\frac{1}{2M\nu}\delta(x-x_i)$ and
%$\delta[(k+q)^2]={\displaystyle\frac{1}{2M\nu}}\delta(x-x_i)$ and
the hadronic tensor becomes
\begin{eqnarray}
W_{\mu\nu}&=&\sum_{i}e^2_i\sum_{s}\frac{d^4k}
{2M\nu}f^i_s(p,k)\delta(x-x_i)\cdot \nonumber \\
&&[2k_{\mu}k_{\nu}+k_{\mu}q_{\nu}+k_{\nu}q_{\mu}-g_{\mu\nu}k.q], \label{37}
\end{eqnarray}
where $e_i$ is the charge of the quark $i$.
From this expression, one can extract $W_{1,2}$ whose features
correspond to the scattering of point-like objects (see eq.(\ref{10}))
and by comparing with eq.(\ref{7}), one finds after some partial
integration and using eq.(\ref{11})
\begin{equation}\label{38}
F_1(x)=\frac{1}{2}\sum_{i} e^2_i q_i(x)\ ,
\end{equation}
where
\begin{equation}\label{39}
q_i(x)=\sum_{s=\pm}\int d^2k_T f^i_s(p,k)=q_{i+}(x)+q_{i-}(x).
\end{equation}
Clearly $q_i(x)$, which is called the {\it unpolarized} quark distribution 
function, contains the sum over two quark spin directions and
the average over the quark intrinsic transverse momentum $k_T$. Usually in the
QPM one interprets $q_{i+(-)}(x)dx$ as the probability to find a quark
$i$, with helicity parallel (antiparallel) to the nucleon spin, and
with momentum fraction of the proton between $x$ and $x+dx$. From
eq.(\ref{37}) one also finds $\nu W_2=2xMW_1$, that is the Callan-Gross
relation (see eq.(\ref{17})) which leads to $R=0$, as mentioned before.
This is what one expects for quarks of spin $1/2$ and let us recall the
argument. Consider a collision quark-virtual photon in the Breit frame.
By conservation of the total spin along the momentum direction, a
scalar quark $(s=0)$ cannot absorb a photon of helicity $\lambda=\pm
1$, so $\sigma_T=0$ and $R=\infty$. However for $s=1/2$, the quark
helicity flips in the electromagnetic interaction and therefore only
the photon with $\lambda=\pm 1$ can be absorbed, so $\sigma_L=0$ and
$R=0$.

Finally let us consider polarized DIS and the interpretation of the
polarized structure functions $g_{1,2}(x)$. Similarly in the QPM, the
antisymmetric part of the hadronic tensor (see eq.(\ref{26})) can be
expressed as
\begin{equation}\label{40}
W_{\mu\nu}^{(A)}\!\!=\!\!\sum_{i}\!\sum_{s} d^4kf^i_s(p,k,s_p) 
w_{\mu\nu}^{(A)i}(q,k,s)\delta[(k+q)^2] ,
\end{equation}
where $s_p$ denotes the proton polarization vector and
\begin{equation}\label{41}
w_{\mu\nu}^{(A)i}=\frac{e^2_i}{4}Tr\left\{(1+\rlap/s\gamma_5) \rlap/k
\gamma_{\mu}(\rlap/k+\rlap/q)\gamma_{\nu}\right\}\ .
\end{equation}
So
\begin{eqnarray}
W_{\mu\nu}^{(A)}&=&i\varepsilon_{\mu\nu\rho
\sigma}q_{\rho}\sum_{i}e^2_i\int d^4k\left[f^i_+(p,k,s_p)-
f^i_-(p,k,s_p)\right]\cdot \nonumber \\
&&k_{\sigma}\delta\left[(k+q)^2\right]\ . \label{42}
\end{eqnarray}
By comparing with eq.(\ref{26}) and using eq.(\ref{30}) one finds
\begin{equation}\label{43}
g_1(x)=\frac{1}{2}\sum_{i} e^2_i\Delta q_i(x)\quad\hbox{and}\quad
g_2(x)=0\ ,
\end{equation}
where with the help of the definitions in eq.(\ref{39})
\begin{equation}\label{44}
\Delta q_i(x)=q_{i+}(x)-q_{i-}(x)\ ,
\end{equation}
which is called the {\it polarized} quark distribution function.
Therefore in the scaling limit the longitudinal polarization of the
proton is described by $g_1(x)$ and we find that $g_2(x)$ vanishes.

Actually if one goes beyond the most naive QPM (see section~\ref{audelaqpm}), 
one must also consider the existence of gluons which produce quark-antiquark
pairs. Since the antiquark contribution can be computed similarly, all
the above quark distributions must be replaced by adding quarks and
antiquarks, so the correct  expressions are indeed
\begin{eqnarray}
&&F_1(x)=\frac{1}{2}\sum_{i}e^2_i\left[q_i(x)+\bar q_i(x)\right]\quad
\hbox{and}\nonumber \\
&&g_1(x)=\frac{1}{2}\sum_{i}e^2_i\left[\Delta
q_i(x)+\Delta \bar q_i(x)\right]\ . \label{45}
\end{eqnarray}
One sees from these expressions that by measuring only $F_1(x)$ (or $F_2(x)$), 
it will not be possible to disentangle neither unpolarized quark distributions
of different flavors, nor quarks from antiquarks. For polarized 
quark (antiquark) distributions the situation is the same if one measures 
only $g_1(x)$. So we give now a short discussion to clarify this point.

Let us go back to $\nu(\bar{\nu})p\to\mu^{\mp}X$ which are interpreted
in the QPM in terms of four different scattering processes, 
{\it i.e.} $\nu q$, $\nu\bar q$, $\bar{\nu}q$ and
$\bar{\nu}\bar q$. For $\nu q$ scattering ({\it resp.} 
$\bar{\nu}\bar q$) both $\nu$ and $q$ are left-handed ({\it resp.}
both $\bar{\nu}$ and $\bar{q}$ are right-handed), so the cross section
$d\sigma/dy$ is isotropic ($s$ wave) whereas for $\nu\bar q$ since $\bar q$ 
is right-handed, in this case ($p$ wave)
%$d\sigma/dy\sim\left({\displaystyle\frac{1+\cos\theta}{2}}
$d\sigma/dy\sim\left(\frac{1+\cos\theta}{2} \right)^2\sim (1-y)^2$, 
({\it idem} for $\bar{\nu}q$ since $q$ is left-handed). 
Therefore one can write for a given flavor
\begin{eqnarray}
&&\frac{d^2\sigma^{\nu}}{dxdy}\sim\left[q(x)+(1-y)^2\bar q(x)\right]\quad
\hbox{and}\nonumber \\
&&\frac{d^2\sigma^{\bar{\nu}}}{dxdy}\sim\left[\bar q(x)+(1-y)^2
q(x)\right]\ . \label{46}
\end{eqnarray}

Consequently the $y$ dependence allows to separate $q(x)$ from $\bar
q(x)$ and by measuring both $\nu$ and $\bar{\nu}$ cross sections, one can
isolate $q(x) - \bar q(x)$ and therefore we have 
(see eqs.(\ref{21}) and (\ref{23}))
\begin{equation}\label{47}
xF^{\nu}_3(x)=\sum_{i} x(q_i(x)-\bar q_i(x))\ .
\end{equation}
A similar analysis can be done to separate $\Delta q(x)$ from 
$\Delta \bar q(x)$ by using charged current reactions in $e^{\pm}p$
polarized DIS, which might be done at HERA DESY in the future.

%%%%%%%%%%%%%%%%%%%%%%%%%%%%%%%%%%%%%%%%%%%%%%%%%%%%%%%%%%%%%%%%%%%%%%%%%%%%%%
\subsection{Main features of quark distributions from DIS data}
%\label{mainfeat}

\begin{figure}[htbp]
\epsfxsize=7.0cm
\centerline{\epsfbox{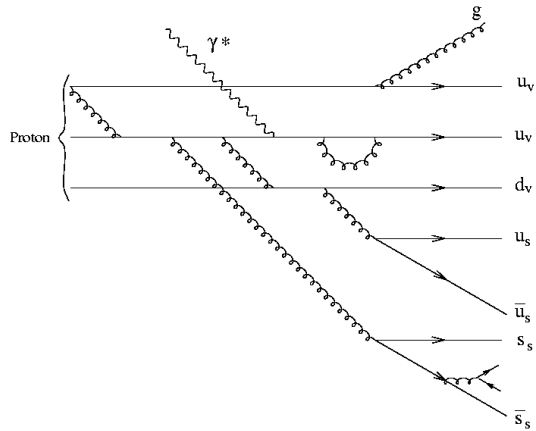}}
\caption{A microscopic view of the proton seen by a $\gamma^*$ having a 
large $Q^2$ and a large $\nu$.}
\label{fig:proton}
\end{figure}

\begin{figure}[ht]
\epsfxsize=7.0cm
\centerline{\epsfbox{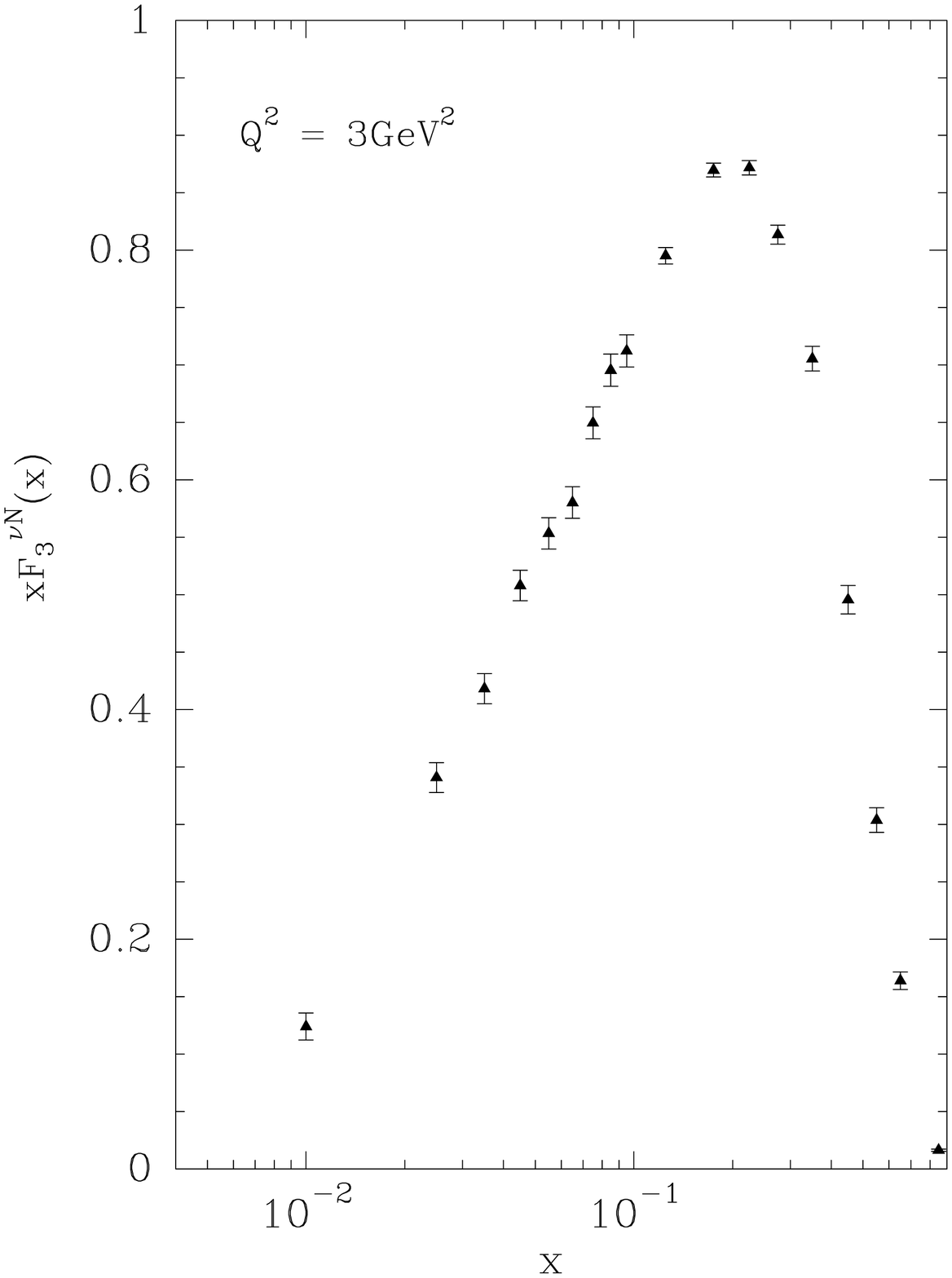}}
\caption{The structure function $xF_3^{\nu N}(x)$ versus $x$ at $Q^2=3GeV^2$
taken from ref.\protect\cite{QUI}.}
\label{fig:xf3x}
%\end{figure}
%%%%%%%%%
%%%%%%%%%%%%%
%\begin{figure}[hb]
\epsfxsize=7.0cm
\centerline{\epsfbox{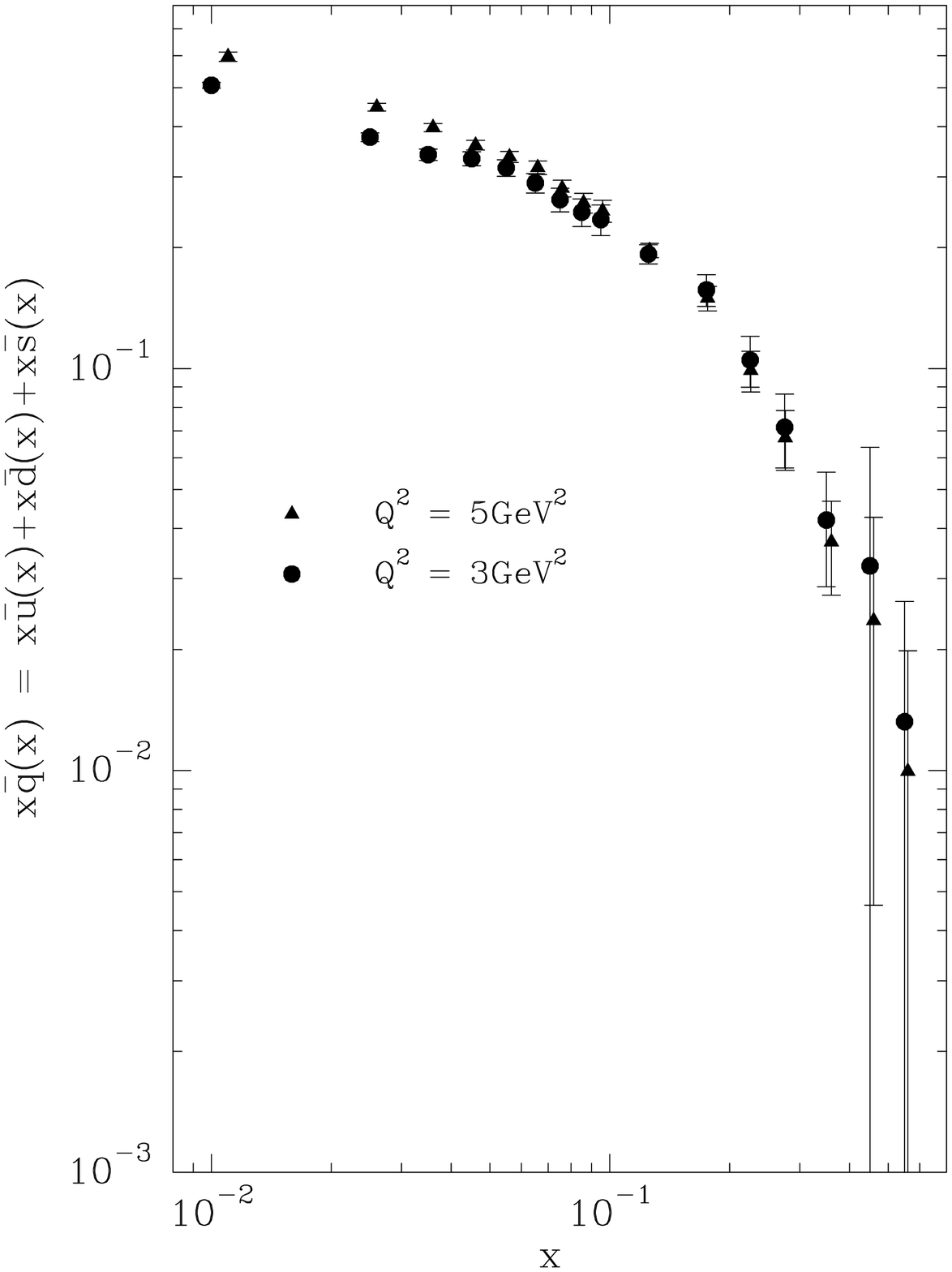}}
\caption{The antiquark distribution at $Q^2=3GeV^2$ (full circles) and
$Q^2=5GeV^2$ (full triangles) versus $x$ taken from ref.
\protect\cite{QUI}.}
\label{fig:qbarx}
\end{figure}

Once the parton structure of the proton has been detected, to establish
it firmly, we need to make a detailed study of its basic constituents. 
A very high energy proton
is not a simple three quarks objects like a proton at rest, but much more 
complex and it is actually made of valence quarks, 
gluons and quark-antiquark pairs which carry a small fraction of its momentum
(see Fig. \ref{fig:proton}).
We define the {\it valence} quark distributions as
\begin{equation}\label{48}
q_{val}(x)=q(x)-\bar q(x) = q(x)-q_{sea}(x)\ ,
\end{equation}
where $q_{sea}(x)$ denotes the sea quark distribution and we assume
$q_{sea}(x)=\bar q(x)$. From neutrino-nucleon DIS via the charged
current, one can extract valence quarks from the measurement of
$F^{\nu}_3(x)$ (see eq.(\ref{47})) and the most accurate data comes
from CCFR\cite{QUI} shown in Fig. \ref{fig:f3ccfr}. By looking at a
fixed $Q^2$ value, as shown in Fig. \ref{fig:xf3x}, we see from this 
pure valence
%%%%%%%%%%%%%
%%%%%%%%%%%%%

\noindent contribution, that $xu_{val}(x)$ and $xd_{val}(x)$ are maximum around
$x\sim0.2$ or so and they vanish for $x\sim 1$ and $x\sim 0$. 
Since one can also isolate the antiquark distributions, 
the results from CCFR\cite{QUI} show that $x\bar q(x)$ (or
$xq_{sea}(x)$) goes to zero very rapidly for $x\sim 1$ and has a sharp
rise for $x\sim 0$ (see Fig. \ref{fig:qbarx}). The small $x$ kinematic 
region which corresponds to $Q^2$ fixed and $\nu\to \infty$, 
is called the Regge region. A simple Regge analysis of the absorption 
cross sections $\gamma^{\star}p$ leads to
\begin{equation}\label{49}
q_{sea}(x)\build{\longrightarrow}_{x\to 0}^{} 1/x \quad\hbox{and}\quad
q_{val}(x)\build{\longrightarrow}_{x\to 0}^{} 1/\sqrt x\ .
\end{equation}
Clearly the singular behavior shown in Fig.\ref{fig:qbarx} is stronger 
than this prediction and we will come back to this important point. 
From $ep\to e'X$ and $ed\to e'X$ DIS on hydrogen and deuterium targets, 
one can extract the electromagnetic structure functions $F^p_2(x)$ and
$F^n_2(x)$ for proton and neutron, in terms of valence quark and
antiquark distributions.

%%%%%%%%%%%%%
\begin{figure}[ht]
\epsfxsize=9.0cm
\centerline{\epsfbox{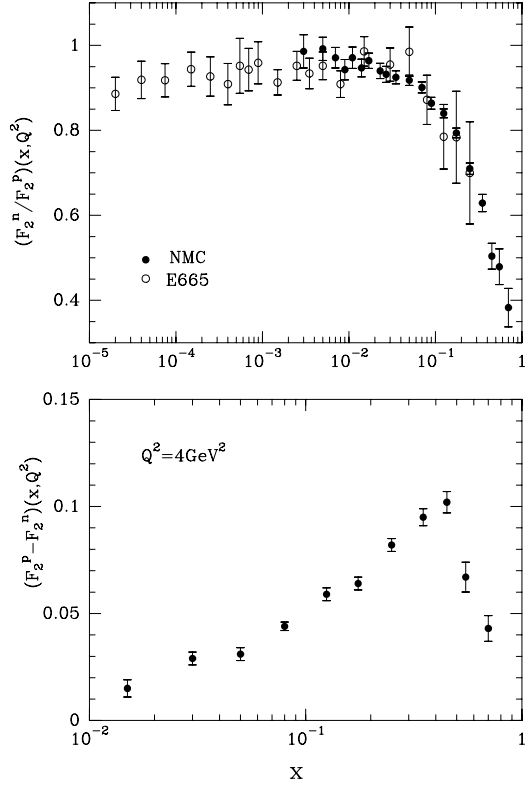}}
\caption{Ratio $F_2^n / F_2^p$ as a function of $x$ for differents $Q^2$
values, data are from NMC\protect\cite{amau95} and E665
\protect\cite{adams95}.
Difference $F_2^p -F_2^n$ as a function of $x$ for $Q^2 = 4\mbox{GeV}^2$,
data are from NMC\protect\cite{ARN}.}
\label{fig:f2pn}
\end{figure}

\begin{figure}[ht]
\epsfxsize=7.0cm
\centerline{\epsfbox{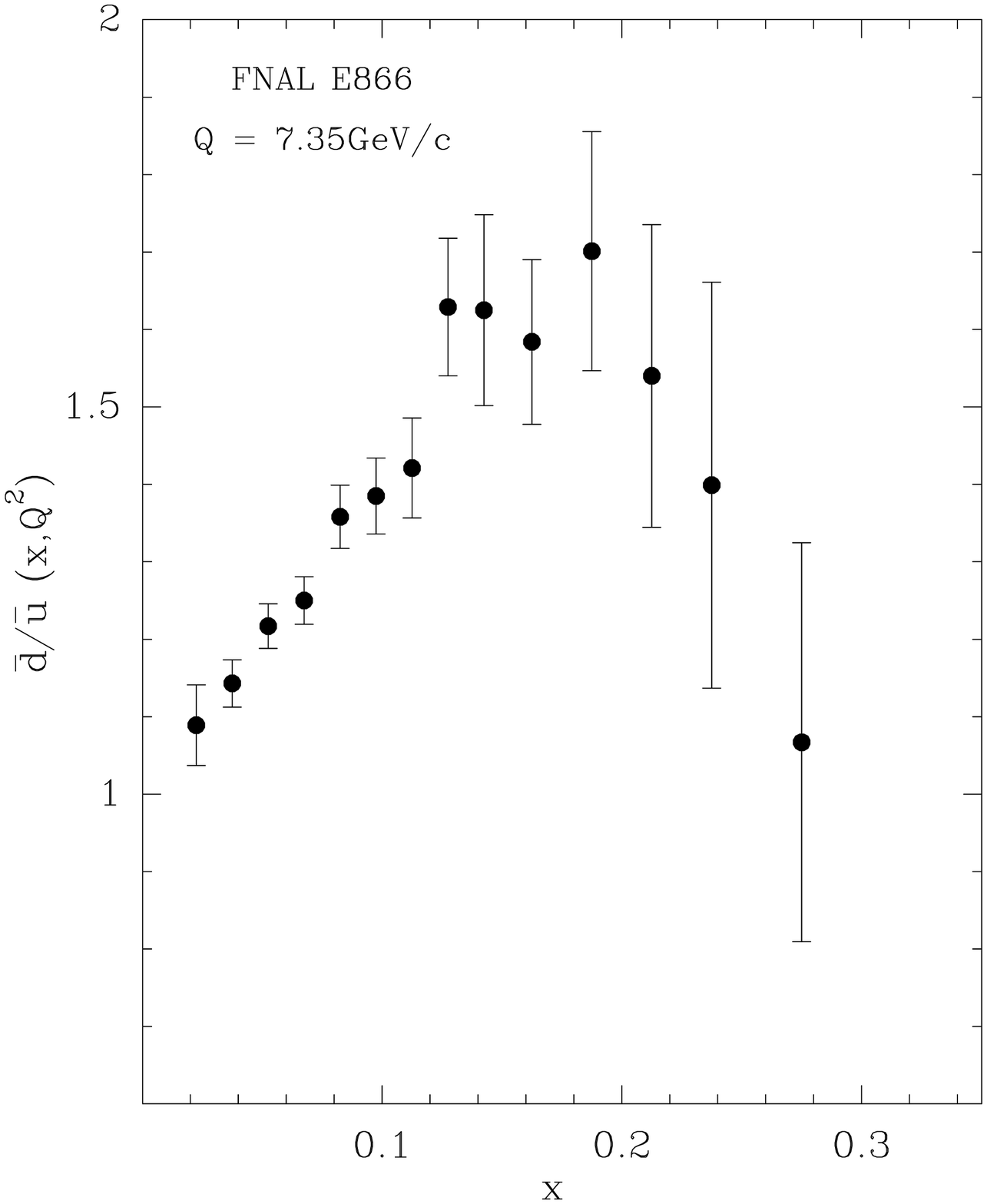}}
\caption{ $\bar d$/$\bar u$ at $Q = 7.35\mbox{GeV/c}$ measured by
FNAL E866 experiment\protect\cite{E866}.}
\label{fig:se866}
%\end{figure}
%%%%%%%%%
%%%%%%%%%%%%%
%\begin{figure}[ht]
\epsfxsize=7.0cm
\centerline{\epsfbox{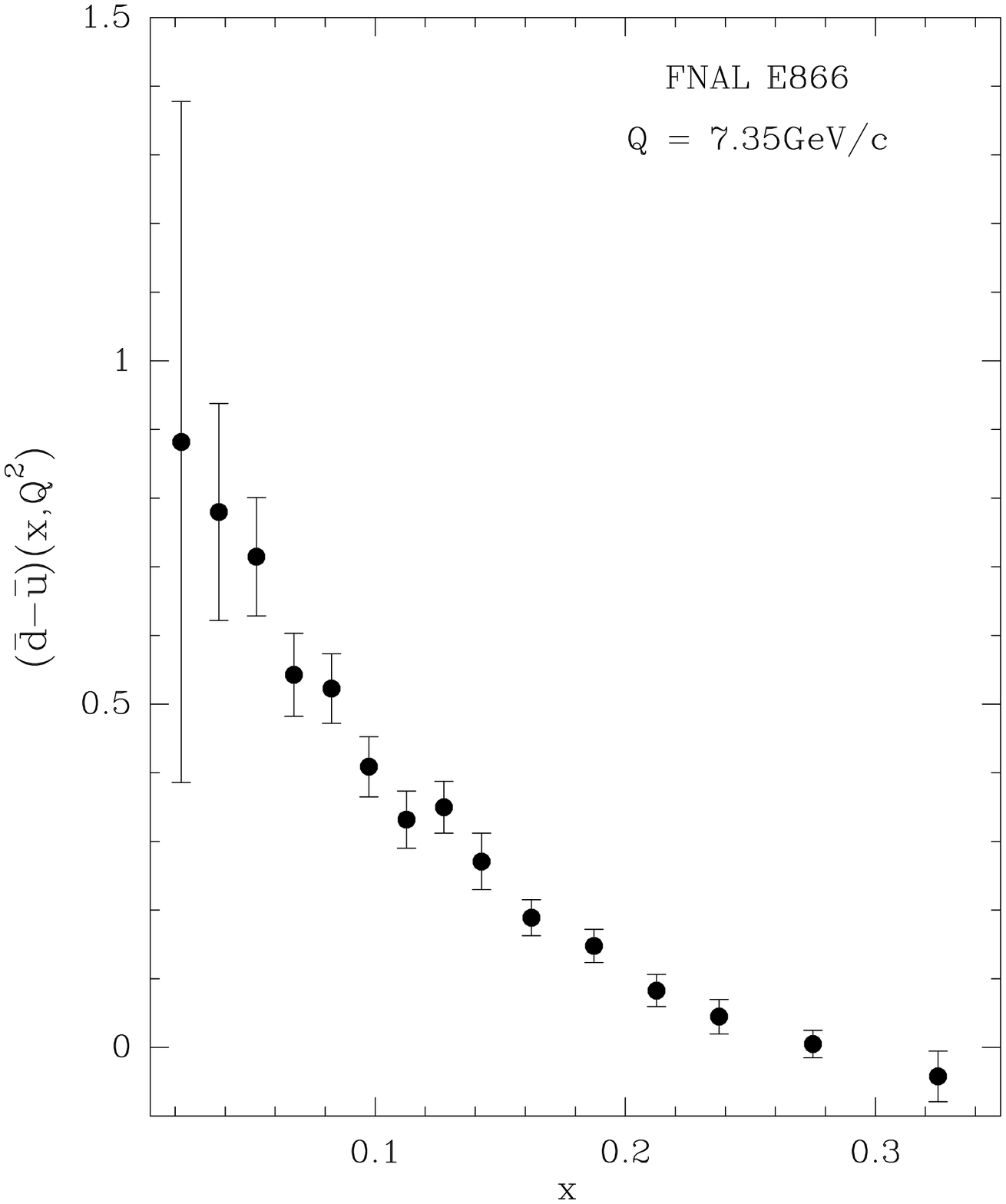}}
\caption{ $\bar d - \bar u$ at $Q = 7.35\mbox{GeV/c}$ extracted from FNAL 
E866 experiment \protect\cite{E866}.}
\label{fig:me866}
\end{figure}

%%%%%%%%%
\clearpage
By using eq.(\ref{17}) and eq.(\ref{45}) we have 
\begin{eqnarray}
F^p_2(x) &=& \frac{4}{9}\left[xu_{val}(x) + 2x\bar u(x)\right]
+\frac{1}{9}\left[xd_{val}(x) + 2x\bar d(x)\right] \nonumber \\
&&+\frac{1}{9}\left[xs(x) + x\bar s(x)\right] \ , \label{50}
\end{eqnarray}
if we restrict ourselves to three flavors $u,d$ and $s$. $F^n_2(x)$ is
obtained from eq.(\ref{50}) by exchanging $u$ and $d$. 
The NMC experiment at CERN\cite{ARN} provides a rather accurate determination
of $F^n_2(x)/F^p_2(x)$ and $F^p_2(x)-F^n_2(x)$ as shown in Fig.\ref{fig:f2pn}.
These data confirm and complete some of the features observed in
neutrino data. The difference $F^p_2(x)-F^n_2(x)$ has a maximum around
$x=0.3$ or so, where $u_{val}(x)$ and $d_{val}(x)$ dominate and it
vanishes for $x\sim 1$. As we will see later, its behavior in the small
$x$ region has revealed the flavor symmetry breaking of the sea quarks,
{\it i.e.} $u_{sea}(x)\not = d_{sea}(x)$.
%%%%%%%%%%%%%
%%%%%%%%%
Since sea quarks dominate for small $x$, we expect the ratio 
$F^n_2(x)/F^p_2(x)$ to go to one in the low $x$ region as indicated
by the NMC data. The E665 data, which have larger errors, suggest a
departure from this trend in the very small $x$ region. 
The behavior of $F^n_2(x)/F^p_2(x)$ for large $x$ 
indicates that $u_{val}(x)$ dominates over $d_{val}(x)$.
The NuSea-experiment\cite{E866}, has measured the ratio of muon pair
yields from Drell-Yan production of proton-proton or proton-deuteron
interactions. This ratio allows the determination of $\bar d /\bar u$
ratio (see Fig. \ref{fig:se866}), which has a strong $x$ dependence
and confirms the antiquark flavor asymmetry, mentioned above and the fact
that $\bar d > \bar u$. Using a parametrization of $\bar d + \bar u$, 
one can also extract $\bar d - \bar u$ (see Fig.\ref{fig:me866}), 
which is mainly positive (except may be at large $x$) and 
exhibits a rapid rise for low $x$, in agreement with what was observed
in neutrino DIS (see Fig. \ref{fig:qbarx}). 
%%%%%%%%%%%%%
%%%%%%%%%

The simple parametrization $\bar d(x) - \bar u(x) = (1-x)^7$,
which was proposed ten years ago \cite{PRS91}, is consistent with these 
new data (see Fig.\ref{fig:me866}). The antiquark flavor asymmetry
is an important topic which has generated new developments and for 
a complete review see ref.\cite{KUM}.

\begin{figure}[hb]
\epsfxsize=8.5cm
\centerline{\epsfbox{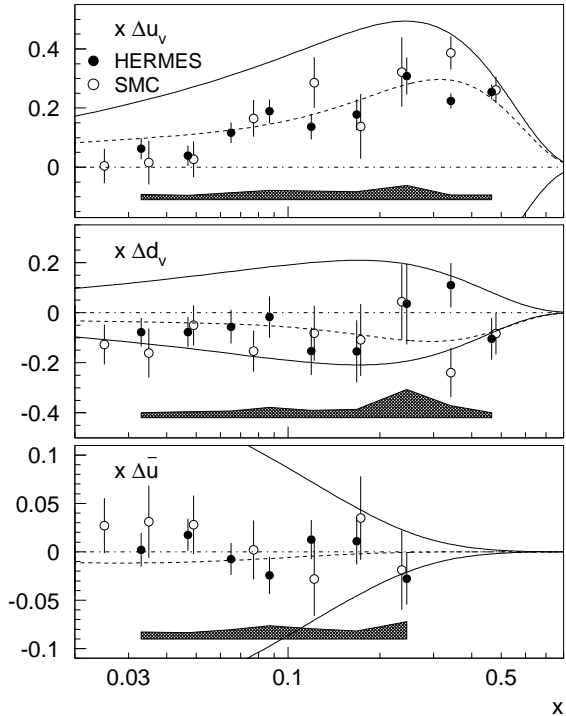}}
\caption{The polarized quark distributions at $Q^2=2.5GeV^2$ separately
for the valence quarks $x\Delta u_v(x)$, $x\Delta d_v(x)$ and the sea
quarks $x\Delta \bar u_v(x)$ as a function of $x$ (Taken from 
ref. \protect\cite{hermes99}).
The solid lines indicate the positivity limit and the dashed lines are 
a model parametrization.}
\label{fig:deltaq}
\end{figure}

We have already said that a direct flavor decomposition of the polarized
quark distributions is not possible
from the data on $g^p_1(x)$ and $g^d_1(x)$ (see eq.(\ref{45})), but this can be
done using polarized semi-inclusive DIS, where one detects a hadron in the final
state. We show in Fig. \ref{fig:deltaq} some recent results from HERMES 
\cite{hermes99} 
for the determination of $x\Delta u_v(x)$, which turns out to be positive, 
$x\Delta d_v(x)$ which is negative and $x\Delta\bar u(x)$ which is compatible 
with zero. The accuracy of these results is limited, 
but these trends agree with some global 
fits of the $g_1^{p,n,d}(x,Q^2)$ data, as we will see later.

We now turn to a short review of different sum rules, the structure
functions we have introduced so far must satisfy. We will also discuss 
their validity or breakdown versus experiment, as well as some 
theoretical considerations. 
%%%%%%%%%%%%%
%%%%%%%%%
%%%%%%%%%%%%%%%%%%%%%%%%%%%%%%%%%%%%%%%%%%%%%%%%%%%%%%%%%%%%%%%%%%%%%%%%%%%%%%%
\subsection{Sum rules}
%\label{sumrul}

There exists a number of sum rules for unpolarized and polarized
structure functions, some of which are rigorous results and other which rely
on more or less well justified assumptions. Let us first consider the
unpolarized case and for the electromagnetic structure functions we
start with the {\it Gottfried sum rule}\cite{GOT} (GSR). If one assumes
an $SU(2)$ symmetric sea, {\it i.e.} $\bar u(x)=\bar d(x)$, one can
easily show, using eq.(\ref{50}), that
\begin{equation}\label{51}
I_G=\int^{1}_{0}\frac{dx}{x} \left[F^p_2(x) - F^n_2(x)\right] =
\frac{1}{3}\ ,
\end{equation}
because
\begin{equation}\label{52}
N_u=\int^{1}_{0}dx\ u_{val}(x)=2\quad\hbox{and} \quad
N_d=\int^{1}_{0}dx\ d_{val}(x)=1\ .
\end{equation}
In fact the NMC experiment\cite{ARN} has observed a large defect of the
GSR, since their measurement (see Fig. \ref{fig:f2pn}) gives at 
$Q^2=4 \mbox{GeV}^2$, $I_G=0.235\pm 0.026$. 
This flavor symmetry breaking, more precisely
$\bar d>\bar u$, is a consequence of the Pauli exclusion principle
which favors $d\bar d$ pairs with respect to $u\bar u$ pairs, since the
proton contains two $u$ quarks and only one $d$ quark. 

Next  for the charged current structure functions, we have two 
rigourous results based on eq.(\ref{52}), namely the 
{\it Adler sum rule}\cite{ADL} (ASR)
\begin{equation}\label{53}
\int^{1}_{0}\frac{dx}{2x} \left[F^{\bar{\nu}p}_2(x) - F^{\nu
p}_2(x)\right] =
N_u-N_d=1\,
\end{equation}
(an experimental verification\cite{ALLA} with large errors show an agreement)
and the {\it Gross-Llewellyn Smith sum rule}\cite{GRO} (GLSSR)
\begin{equation}\label{54}
I_{GLS}\!\!=\!\!\int^{1}_{0}\frac{dx}{2x} \left[xF^{\nu p}_3(x) + x
F^{\bar{\nu} p}_3(x)\right] \!\!=\!\!
N_u+N_d\!= \!3 .
\end{equation}
The ASR is exact and receives no QCD corrections, but its experimental
verification is at a very low level of accuracy \cite{ALLA}. The GLSSR gets
a negative QCD correction and the CCFR data (see Fig.\ref{fig:f3ccfr}) 
gives\cite{QUI}
$I_{GLS}=2.55\pm 0.06\pm 0.1$ at $Q^2=3\mbox{GeV}^2$, in fair agreement with
the theoretical prediction.
If we turn to polarized structure functions, there is first a
fundamental result called the {\it Bjorken sum rule}\cite{BJOR} (BSR).
It was derived about thirty years ago in the framework of quark current
algebra and it relates the first moment of the difference between
$g^p_1(x)$ for the proton and $g^n_1(x)$ for the neutron and the
neutron $\beta$-decay axial coupling
\begin{equation}\label{55}
\int^{1}_{0}dx\left[g^p_1(x) - g^n_1(x)\right] =\frac{1}{6} g_A/g_V\ ,
\end{equation}
where $g_A/g_V = 1.2573\pm 0.0028$ is very accurately known. The BSR
gets also a negative QCD correction and we will come back later to the
test of this firm prediction of QCD. One can also derive sum rules for
$g^p_1$ and $g^n_1$ separately. 
These are the {\it Ellis-Jaffe sum rules}\cite{ELL} (EJSR) which read
\begin{eqnarray}
&&\Gamma_1^p= \int^{1}_{0}dx g^p_1(x)=\frac{1}{18}
(9F-D+6\Delta s)\quad\hbox{and}
\nonumber \\
&&\Gamma_1^n=\int^{1}_{0}dx g^n_1(x)=\frac{1}{18} (6F-4D+6\Delta s)\ ,
\label{56}
\end{eqnarray}
where $F=0.459 \pm 0.008$ and $D=0.798 \pm 0.008$ are the $\beta$-decay axial 
coupling constants of the
baryon octet and $\Delta s= \int^{1}_{0}\Delta s(x)dx$ is the total
polarization of the proton carried by the strange quarks. One recovers
eq.(\ref{55}) by taking the difference because $F+D= g_A/g_V$. 
In their original work, Ellis and Jaffe made the
critical assumption that $\Delta s=0$, which allows to make definite
predictions for $\Gamma^p_1$ and $\Gamma^n_1$. 
%%%%%%%%%%%%%
\begin{figure}[htbp]
\epsfxsize=8.5cm
\centerline{\epsfbox{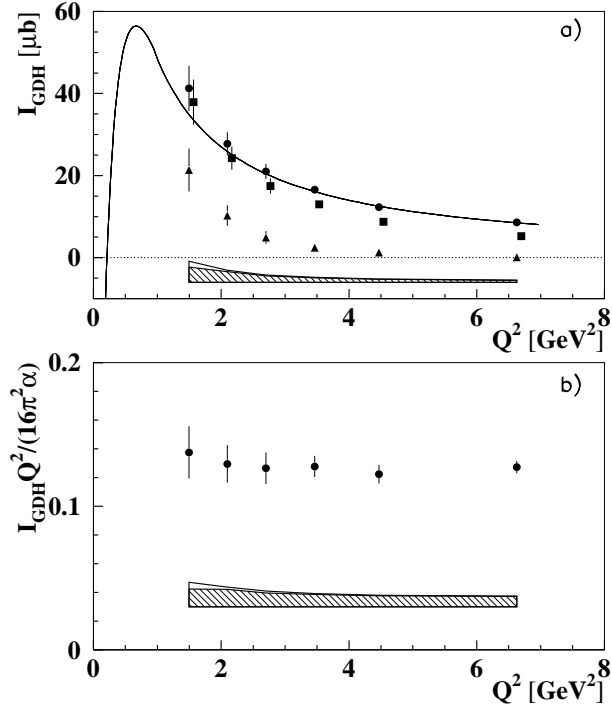}}
\caption{a) $I_{GDH}$ for proton, as a function of $Q^2$ for various upper 
limits of
integration $W^2 \leq 4.2 \mbox{GeV}^2$ (triangles), $W^2 \leq 45 \mbox{GeV}^2$
(squares), the total integral (circles) from HERMES col.\protect\cite{HERMES2}.
The curve is a theoretical prediction\protect\cite{softer2}. b) 
$I_{GDH} Q^2/(16\pi^2\alpha)$ as a function of $Q^2$ shows the weak 
$Q^2$-dependence of $\Gamma_1$.}
\label{fig:igdh}
\end{figure}
%%%%%%%%%
Using the spin-dependent photoabsorption cross sections 
$\sigma_{1/2(3/2)}(\nu)$, {\it Gerasimov-Drell-Hearn}\cite{GERA} (GDH) 
have derived the following sum rule valid for real photons
\begin{equation}\label{eq:gdh}
I_{GDH}= \int_{\nu_{thr}}^\infty [\sigma_{1/2}(\nu)-\sigma_{3/2}(\nu)] 
\frac{d\nu}{\nu} = - {2\pi^2\alpha \over M^2}\kappa^2 \ ,
\end{equation}
where $\nu$ is the photon energy in the target rest frame,
$\nu_{thr}$ is the pion production threshold and $\kappa$ the anomalous nucleon
magnetic moment.
This important sum rule which predicts $-204 \mu b$ for the proton, has never 
been tested
directly due to the need for a circularly polarized beam with a longitudinally 
polarized target and a wide range of photon energies to be covered.
However the GDH integral can be generalized to the case of absorption of
polarized transverse virtual photons with $Q^2$
\begin{equation}\label{eq:ggdh}
I_{GDH}(Q^2)=\int_{\nu_{thr}}^\infty[\sigma_{1/2}(\nu,Q^2)-
\sigma_{3/2}(\nu,Q^2)]\frac{d\nu}{\nu}.
\end{equation}
One can show that, within a good approximation, one has in the scaling limit
\begin{equation}\label{eq:GDH}
I_{GDH}(Q^2)= \frac{16 \pi^2 \alpha}{Q^2} \Gamma_1 \,
\end{equation}
where $\Gamma_1= \int^{1}_{0}dx g_1(x)$, so the GDH sum rule is connected to 
polarized DIS.
The $Q^2$-dependence of the generalized GDH sum rule has been measured by the 
HERMES Collaboration\cite{HERMES2} in the range 
$1.2 \le Q^2 \le 12 \mbox{GeV}^2$ 
(see Fig.\ref{fig:igdh}). The $Q^2$-behavior of $I_{GDH}(Q^2)$ 
agrees with the theoretical prediction of ref.\cite{softer2} and this suggests 
that there are no important effects from either resonances 
or non-leading-twist. 

\begin{figure}[htbp]
\epsfxsize=8.0cm
\centerline{\epsfbox{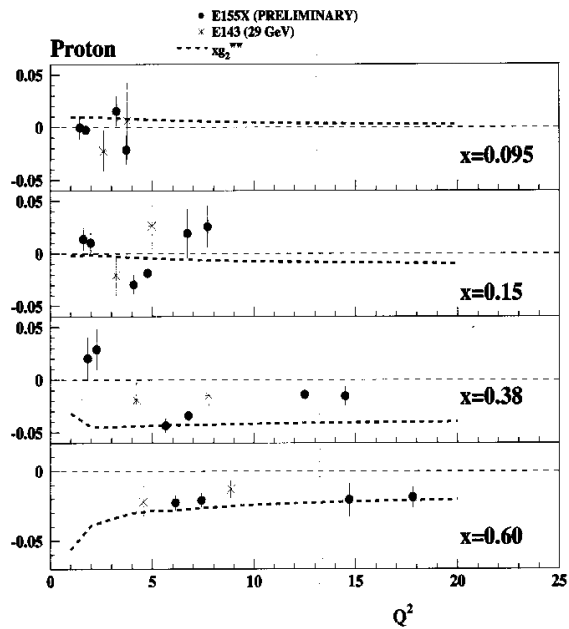}}
\caption{The structure function $x g^p_2$ versus
$Q^2$ for different $x$ values. Preliminary data from SLAC E155X
\protect\cite{E155X}.}
\label{fig:g2e155}
\end{figure}

Concerning the structure function $g_2(x)$, which is related to
transverse polarization (see eq.(\ref{34}), but has no simple interpretation 
in the parton model, it is possible to derive a superconvergence relation by
considering the asymptotic behavior of a particular virtual Compton
helicity amplitude. This leads to the 
{\it Burkhardt-Cottingham sum rule}\cite{BUR} (BCSR)
\begin{equation}\label{57}
\int^{1}_{0}dxg^p_2(x)= 0 \quad\hbox{and}  \quad \int^{1}_{0}dxg^n_2(x)=
0\ ,
\end{equation}
for proton and neutron and from this result, it has been naively argued
that $g_2(x)$ vanishes identically. Actually from what we discussed in
section~\ref{kindis} about the asymmetry $A_2$ in the scaling limit, one can
alternatively expect $g_1(x) +g_2(x) = 0$, a simple relation between
$g_1$ and $g_2$. However $g_2(x)$ is more complicated than
that\cite{JI} and only part of it (twist-$2$ contribution) is entirely
related to $g_1(x)$ by means of the {\it Wandzura-Wilczek sum
rule}\cite{WAN} (WWSR) which reads for $J\geq 1$
\begin{equation}\label{58}
\int^{1}_{0}dx x^{J-1}\left[\frac{J-1}{J} g_1(x) + g^{WW}_2(x)\right] =
0\ .
\end{equation}
Clearly for $J=1$ one recovers the BCSR eq.(\ref{57}) and for $J=2$ one
has
\begin{equation}\label{59}
g^{WW}_2(x)= \int^{1}_{x} g_1(y) \frac{dy}{y} - g_1(x)\ .
\end{equation}
Very preliminary data obtained recently by E155X\cite{E155X} and shown 
in Fig.\ref{fig:g2e155} for proton, are consistent with eq.(\ref{59}), as
shown by the dashed curves, 
but this has to wait for further confirmation. This approximation gives also 
the same kind of agreement with the $A_2$ data for proton and deuteron, 
as noticed above in Fig.\ref{fig:a2e155}.
%%%%%%%%%%%%%
%%%%%%%%%

%%%%%%%%%%%%%%%%%%%%%%%%%%%%%%%%%%%%%%%%%%%%%%%%%%%%%%%%%%%%%%%%%%%%%%%%%%
\section{Beyond the QPM, scaling violations and phenomenological tests}
\label{audelaqpm}

%%%%%%%%%%%%%%%%%%%%%%%%%%%%%%%%%%%%%%%%%%%%%%%%%%%%%%%%%%%%%%%%%%%%%%%%%%%
\subsection{Scaling violations}
%\label{scalviol}

So far we have considered the proton structure in the framework of the
QPM, a picture valid only in the scaling limit. For finite $Q^2$, there
are several experimental facts which confirm the existence of scaling
violations and we have already shown in Figs. \ref{fig:f2p} and 
\ref{fig:f3ccfr}, the $Q^2$ dependence of $F_2^p(x,Q^2)$ and 
$xF^{\nu N}_3(x,Q^2)$, for different values of $x$. For small $x$,
$F_2^p(x,Q^2)$ increases with larger $Q^2$ and for large $x$, it
decreases with larger $Q^2$. Actually this can be understood in the
framework of QCD, a color gauge theory, which is the underlying field
theory for describing quark interactions and in particular, deep
inelastic phenomena.

From the knowledge of the structure functions
$F^p_2, F^n_2, F^{\nu p}_2$ and $F^{\nu n}_2$, one basic observation,
which was made long time ago, is the fact that in DIS, the total
fraction of the proton momentum carried by quarks and antiquarks, {\it
i.e.} $\sum_{j}^{}\int^{1}_{0} dx\left[xq_j(x)+x\bar q_j(x)\right]$, is
about $1/2$. So half of the proton momentum is carried by electrically
neutral objects, which have been identified with the gluons, already
mentioned above. Gluons are the carriers of the color force and
therefore in the QPM diagram shown in Fig.\ref{fig:graph2} 
(see also Fig.\ref{fig:proton}), one can 
no longer neglect the possibility that the quark may radiate a gluon, 
before or after being struck by the virtual photon $\gamma^{\star}$. 
These contributions are of order $\alpha\alpha_s(Q^2)$, where $\alpha_s(Q^2)$
is the strong interaction coupling and moreover a gluon as a parton in
the proton can contribute to DIS via the diagram $\gamma^{\star}g\to
q\bar q$ to the same order. From the QCD Lagrangian and the
corresponding Feynman rules one can compute these contributions, so
partons in QCD\cite{ALT} are no longer described by distributions which
are functions of $x$ alone, but of $x$ as well as $Q^2$. 
These scaling violations will affect the predictions of the QPM and, as 
an example, the ratio $R$ (see eq.(\ref{16})) is not equal to zero, so the
Callan-Gross relation (see eq.(\ref{17})) is not exact any more, and so
on. The $Q^2$ dependence of the parton (quark, antiquark and gluon)
distributions has been investigated by several authors\cite{GRI} and
can be summarized in the following way. Given the parton distributions
at some reference point $Q^2=Q^2_0$, the $Q^2$ evolution is given by a
set of coupled equations between singlet part\footnote{There is a
similar equation for the non-singlet $q^{NS}=q-\bar q$ with no gluon
contribution.} $q^S=q+\bar q$ and gluon $G$
\begin{eqnarray}
&&\frac{d}{d\ell nQ^2}\left(\begin{array}{c}q^S(x,Q^2)\\
G(x,Q^2)\end{array}
\right) = \frac{\alpha_s(Q^2)}{2\pi} \cdot\nonumber \\
&&\int^{1}_{x}\frac{dz}{z} \left(
\begin{array}{cc}P_{qq}(z)&N_f P_{qG}(z)\\P_{Gq}(z)&
P_{GG}(z)\end{array}
\right) \left(\begin{array}{c}q^S(x/z,Q^2)\\ G(x/z,Q^2)\end{array}
\right) , \label{60}
\end{eqnarray}
where $N_f$ is the number of flavors and the $P_{ij}(z)$, the so-called 
splitting functions, are known\cite{ALT,GRI}. This set of equations, usually
called the DGLAP equations, is valid in leading order (LO) QCD, but
it can be generalized to next-to-leading order (NLO), {\it i.e.} by including
corrections of order $(\alpha_s(Q^2))^2$, because we have
\begin{equation}\label{Pij}
P_{ij}(x)= P_{ij}^{(0)}(x) + \frac {\alpha_s}{2\pi} P_{ij}^{(1)}(x) +....
\end{equation}
A more formal approach to derive the $Q^2$ behavior of the structure
functions is based on the use of the operator product
expansion\cite{WIL} (OPE). The starting point is to consider time
ordered products of two currents, which are related to the hadronic
tensors introduced above, and to study their singularities on the light
cone\cite{BRA}. This is also a very powerful method to derive directly
sum rules in QCD, together with their radiative corrections at the
level of one or several loops.

Concerning the polarized distributions there exists a set of equations
similar to eq.(\ref{60}) which provides their $Q^2$
evolution\cite{ALT}, and the corresponding polarized splitting functions 
can be found in refs.\cite{WVog,Mertig}. 
We note that in addition to the quark (antiquark)
polarizations $\Delta q_i(x,Q^2)$, one also introduces the gluon
polarization $\Delta G(x,Q^2)$ defined as in eq.(\ref{44}). This
distribution plays also a crucial r\^ole in connection with the
anomaly of the axial current\cite{ALTA}, as we briefly recall
now\cite{ANS}. For a given flavor $i$, the first moment $\Delta q_i$
is defined as the nucleon matrix element of the quark axial current
\begin{equation}\label{61}
<p,s|\bar q_i\gamma_{\mu}\gamma_5q_i |p,s> = 2s_{\mu}\Delta q_i\ .
\end{equation}
Due to the fact that this axial current is not conserved, it was
shown\cite{ALTA} that eq.(\ref{61}) gets a gluon contribution which
generates a shift of $\Delta q_i$
\begin{equation}\label{62}
\Delta q_i\to \Delta q_i - \frac{\alpha_s(Q^2)}{2\pi} \Delta G(Q^2)\ .
\end{equation}
It is worth recalling that from the QCD evolution equations $\Delta
q_i$ is $Q^2$ independent, whereas $\Delta G(Q^2)$ increases
logarithmically with $Q^2$, but such an increase exactly compensates
the decrease of $\alpha_s(Q^2)$ with $Q^2$. However the size of this gluonic
correction, which can be partly absorbed in the definition of the
quark distributions, is purely a matter of scheme convention \cite{chen96}. 
Finally we note that so far, we have no direct experimental information 
on the sign and magnitude of $\Delta G$.

%%%%%%%%%%%%%%%%%%%%%%%%%%%%%%%%%%%%%%%%%%%%%%%%%%%%%%%%%%%%%%%%%%%%%%%%%%%%
\subsection{Phenomenological tests}
%\label{phentest}

There is an enormous literature devoted to phenomenological work on DIS
but we shall discuss only very briefly, a few aspects of this broad
area of high energy physics. Clearly one particular goal is to parametrize
all the parton distributions as function of $x$, at a given low $Q^2$ 
value, $Q^2=Q_0^2$, and to test whether the $Q^2$ evolution agrees with the QCD
predictions. In several analysis\cite{MAR} the quark and gluon
parton densities were described by a polynomial standard form
\begin{equation}\label{pdf1}
xp(x,Q^2_0) = \eta A x^a (1 -x)^b (1+ \gamma x + \rho\sqrt{x}) \ ,
\end{equation}
where $\eta, a, b, \gamma, \rho$ are free parameters. These twenty or 
so parameters were determined by making a global NLO fit of the data on 
unpolarized structure functions and the results obtained are rather 
satisfactory. 

In a different approach, one is using, for valence quarks and antiquarks, 
a set of Fermi-Dirac distributions\cite{BOU}, with the following expression 
\begin{equation}\label{pdf2}
xp(x,Q^2_0) = {A x^a \over \exp[ (x - V(x))/ \bar x] + 1 } \ ,
\end{equation}
where $V(x)$ plays the role of the "thermodynamical potential" and
$\bar x$ is the "temperature". The gluon density function $G(x,Q^2_0)$ is given
in this approach by a Bose-Einstein distribution with no free parameter.

In this case we have a small number of parameters, say ten at most, which were
determined at $Q^2=4 \mbox{GeV}^2$ by fitting the CCFR data displayed in
Fig. \ref{fig:xf3x} for valence quarks, in Fig. \ref{fig:qbarx} for antiquarks 
and the NMC data on the ratio and the difference of $F^p_2$ and $F^n_2$ 
in Fig. \ref{fig:f2pn}.
%%%%%%%%%%%%%
\begin{figure}[htbp]
\epsfxsize=7.5cm
\centerline{\epsfbox{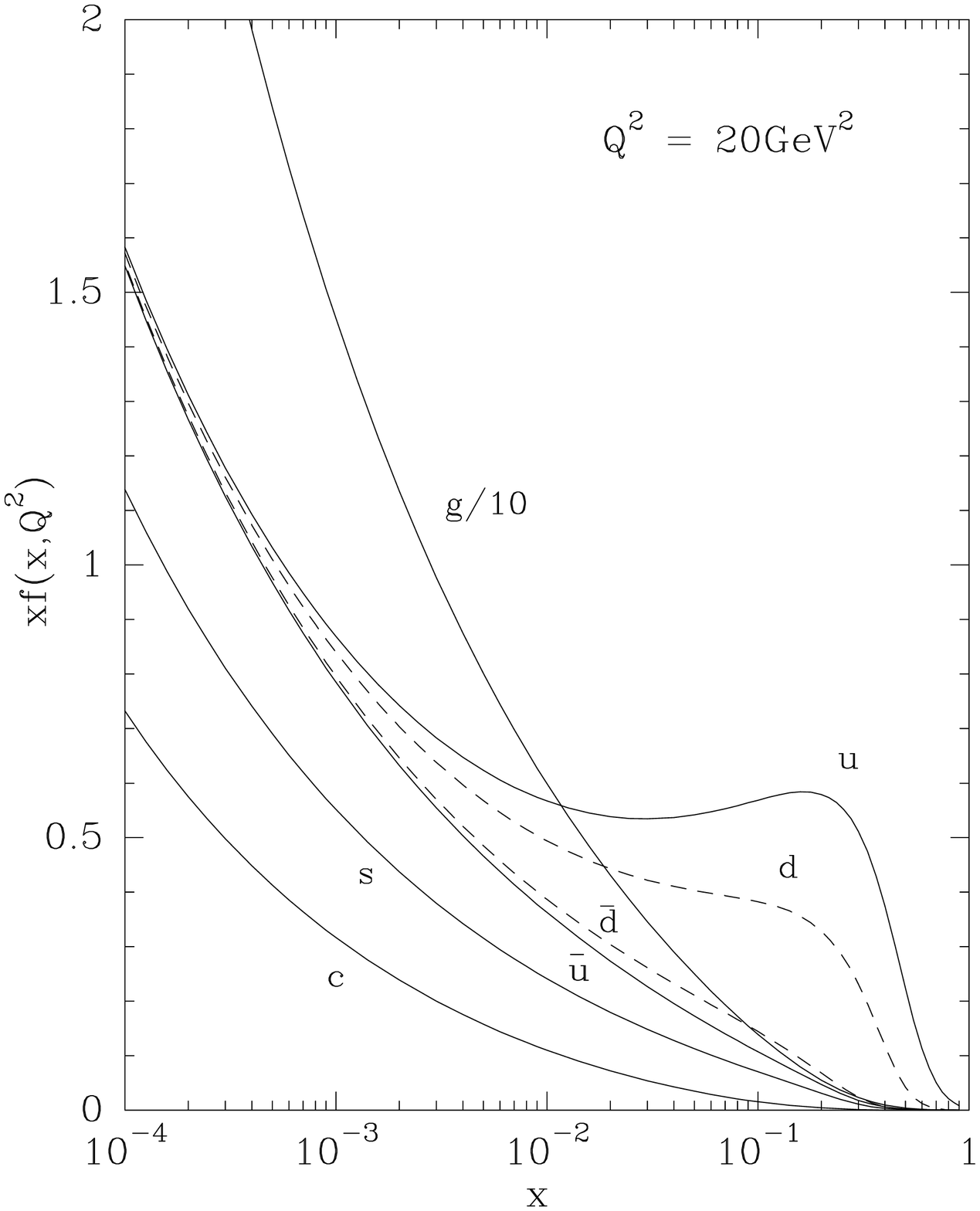}}
\caption{Typical unpolarized parton density functions obtained from
ref.\protect\cite{BOU}.}
\label{fig:uparton}
%\end{figure}
%%%%%%%%%
%%%%%%%%%%%%%
%\begin{figure}[htbp]
\epsfxsize=7.5cm
\centerline{\epsfbox{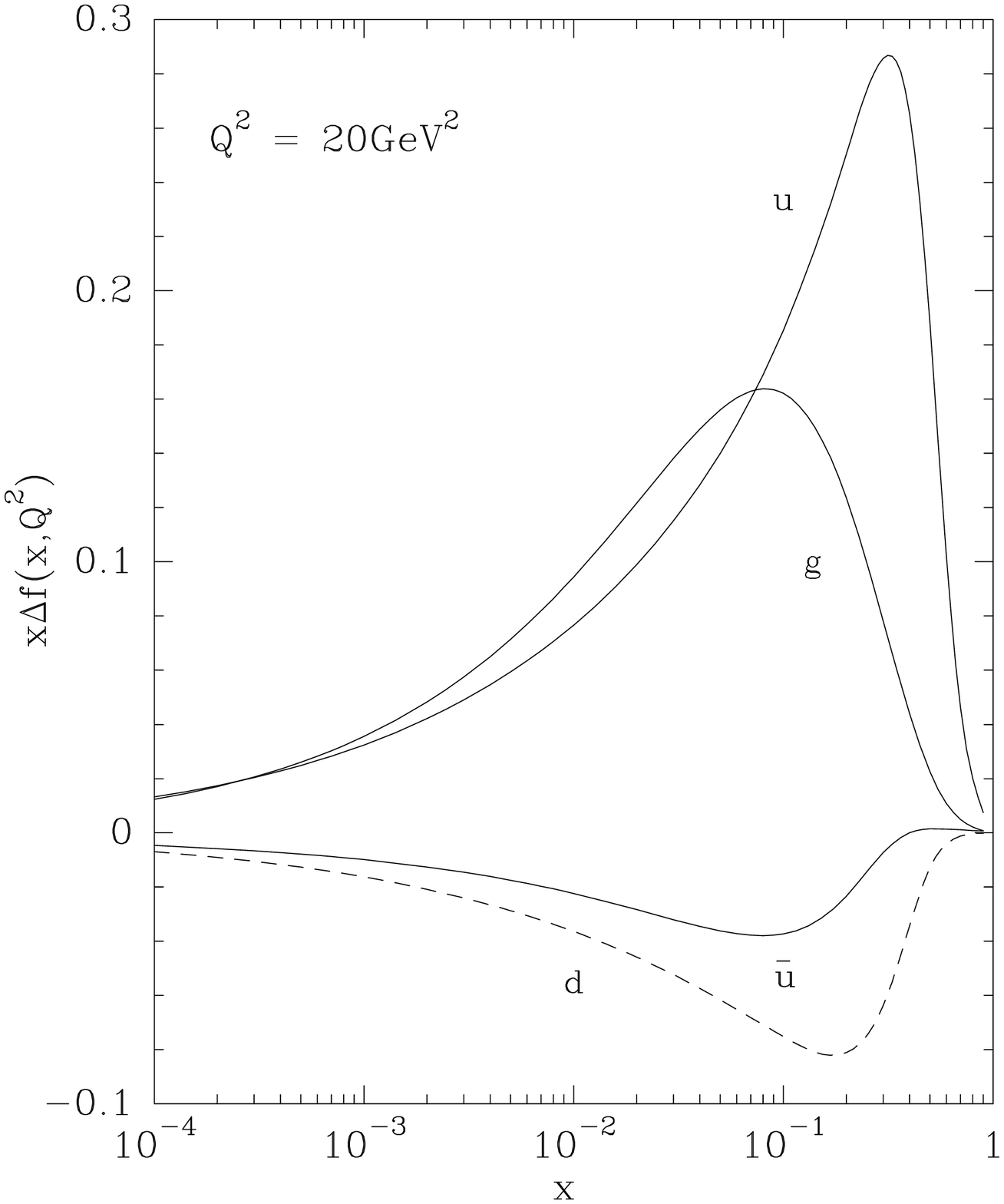}}
\caption{Typical polarized parton density functions obtained from
ref.\protect\cite{BOU}.}
\label{fig:pparton}
\end{figure}
%%%%%%%%%
\clearpage
From these distributions at $Q^2=4 \mbox{GeV}^2$, 
after a standard $Q^2$ evolution using the DGLAP equations at LO only 
\footnote{A new work updating ref.\cite{BOU} with new data and a NLO 
$Q^2$ evolution, is in preparation.}, one obtains the unpolarized 
parton distributions for different flavors, shown in
Fig. \ref{fig:uparton} at $Q^2=20 \mbox{GeV}^2$. We note that, as expected 
$u$ quarks dominate over $d$ quarks, $\bar d(x) > \bar u(x)$ and the $s$ 
and $c$ quarks are very small.
Clearly the gluon distribution $G(x,Q^2)$, which was divided by a factor 
10 and represented by the curve $g/10$ in the Figure, 
dominates largely in the small $x$ region. 
The predictions obtained in a larger $Q^2$ range are in excellent agreement 
with the CCFR data for  $x F_3^{\nu N} (x, Q^2)$ depicted in 
Fig. \ref{fig:f3ccfr} and also with the very recent data at HERA-DESY from the
Zeus and H1 collaborations\cite{DER,AHM} on $F_2^p (x, Q^2)$ displayed
in Fig. \ref{fig:f2p}. 
The rapid rise in the very small $x$ region, down to $x = 10^{- 6}$ 
or so, where sea quarks dominate, is well described by the QCD predictions
and this rise is the direct extrapolation of what has been observed by
CCFR for $x = 10^{- 2}$.
We gave only these two examples for illustration, 
but these tests have been widely extended as discussed in ref.\cite{BOU}. 
The large amount of available data on the
unpolarized structure functions was also used to perform a global
analysis by several authors\cite{MAR} and they all find for the parton 
densities a general pattern similar to the one displayed in 
Fig. \ref{fig:uparton}.

We now turn to some tests involving the polarized structure functions.
As we have seen before, there is a substantial experimental information
on the $x$ dependence of $g_1^{p,n} (x)$ with proton, deuterium and $^3 H e$
polarized targets, for $Q^2$ in a rather limited range, {\it i.e.\/}
$2 \le Q^2 \le 10 \mbox{GeV}^2$. Given the present accuracy of the data and 
the small $Q^2$ range investigated so far, the $Q^2$ evolution of 
$g_1^{p,n} (x, Q^2)$ predicted by QCD cannot be tested at the level it was 
done for the unpolarized case. Moreover this test depends on 
$\Delta G (x, Q^2)$, for which only theoretical speculations can be made. 
In ref.\cite{BOU}, since the initial quark (antiquark) distributions are 
$q_{i \pm} (x, Q^2)$ for each flavor $i$, one gets immediately 
$\Delta q_i (x, Q^2)$. Therefore one can calculate consistently the 
$F_2$'s and the $g_1$'s structure functions, with
no additional parameters and the agreement with the
data is rather good\cite{BOU}. We show in Fig. \ref{fig:pparton} a typical 
pattern of polarized parton distributions. $\Delta u(x)$ and $\Delta d(x)$ are
rather well determined and they are consistent with the results shown 
in Fig. \ref{fig:deltaq}. We also see that $\Delta \bar u(x)$ is small and 
we find that $\Delta \bar q(x)$, for $q=d,s,c$ are compatible
with zero. The $\Delta G (x)$ shown in the Figure is one possible solution 
but it has large uncertainties. There are several other QCD analysis of
the $g_1$'s\cite{GEH}, where one relates the polarized parton
distributions to the unpolarized ones with some additional
parameters. These analysis conclude that $\Delta G$ is weakly
constrained by present data and they all agree that a small $\Delta G$
is prefered.

Finally let's say a few words on testing the first moment of $g_1$,
{\it i.e.\/} the EJSR and the BSR. Since data are obtained for each $x$
at a different $Q^2$, in order to have $g_1 (x, Q^2)$ for a constant 
$Q^2$, one has to rely on the crucial hypothesis that the asymmetry
$A_1$ (see eq.(\ref{28})) is $Q^2$ independent, so the $Q^2$ evolution
of $g_1$ is driven by that of $F_2$. 

Moreover, the sum rules involve integrals from $x = 0$ to $x = 1$ 
and since there is no data at both ends, 
one has to make some reliable extrapolations in these two
regions, low $x$ being the most critical one. The QPM results given in
eqs.(\ref{55}) and (\ref{56}) are modified by QCD corrections, which have
been calculated up to order $(\alpha_s (Q^2))^3$. For example, in the gauge
invariant factorization scheme,
perturbative QCD corrections to $\Gamma_1^{p(n)}$ lead to the following 
results\cite{LAV} :
\begin{equation}
\Gamma_1^{p(n)}=\,C_{NS}\left(\pm{1\over 12}g_A^3+{1\over 36}g_A^8
\right)+{1\over 9}\,C_Sg_A^0,
\end{equation}
where
\begin{eqnarray}
g_A^0&=&\Delta\Sigma\equiv\Delta u+\Delta d+\Delta s,
\nonumber \\
g_A^3&=&F + D, \\
g_A^8&=&3F - D, \nonumber
\end{eqnarray}
the  coefficients non-singlet $C_{NS}$ and singlet $C_{S}$ are given by
\begin{eqnarray}
C_{NS} &=& 1-{\alpha_s\over \pi}-{43\over 12}\left({\alpha_s\over\pi}
\right)^2-20.2153\left({\alpha_s\over \pi}\right)^3,  \nonumber \\
C_{S} &=& 1-{\alpha_s\over \pi}-1.10\left({\alpha_s\over\pi}\right)^2 ,
\end{eqnarray}
for three quarks flavors.
The BSR becomes now 
\begin{equation}\label{63}
\Gamma_1^p - \!\Gamma_1^n = \frac{1}{6} \frac{g_A}{g_V}
\left[
1 - \!\frac{\alpha_s}{\pi} - {43\over 12}\! 
\biggl( \frac{\alpha_s}{\pi} \biggr)^2 - 20.2153\! 
\biggl( \frac{\alpha_s}{\pi} \biggr)^3
\right].
\end{equation}
 Here we have neglected the theoretical uncertainties 
 in the very low $Q^2$ region due to the higher-twist effects
which go like $1 / Q^2$.

The world data for the $\Gamma_1 (Q^2)$'s show a defect in the EJSR for 
proton and deuterium both at $Q^2 = 3 \mbox{GeV}^2$ from E143\cite{ABE,ABEal} 
and $Q^2 = 10 \mbox{GeV}^2$ from SMC\cite{SMC,ADA}. From this one deduces 
also a defect for neutron, which was confirmed by the E154 result. 
If one combines the world data one finds for the BSR
\begin{equation}\label{64}
\Gamma_1^p - \Gamma_1^n = 0.171 \pm 0.005 \pm 0.01
\quad \mbox{at} \quad Q^2 = 5 \mbox{GeV}^2\ ,
\end{equation}
to be compared to the QCD prediction $0.182 \pm 0.005$. 
One can also turn the argument around and given the experimental result, 
use the theoretical expression for the BSR to extract 
$\alpha_s (Q^2)$\cite{KAR}. We also note that the defect in
$\Gamma_1^p$ and $\Gamma_1^n$ are usually interpreted as being mainly due
to a large and negative $\Delta s$ (see eq.(\ref{56})), {\it i.e.\/} 
$\Delta s = - 0.13 \pm 0.02$, but of course this does not necessarily
holds if one considers $SU(3)$ symmetry breaking effects\cite{LIC}. 
The total contribution of the quark spins to the proton
spin, is defined as $\Delta \Sigma = \Delta u + \Delta d + \Delta s$.
%presented in Fig.15 at $Q^2 = 5 \mbox{GeV}^2$. 
For $\Delta s = 0$, the Ellis-Jaffe\cite{ELL} prediction gives 
$\Delta \Sigma \simeq 0.58$, but according to
all experiments, one gets a smaller value, for instance,
E155 experiment\cite{E155} has obtained $\Delta \Sigma = 0.23\pm 0.04 \pm 0.6$ 
at $Q^2 = 5 \mbox{GeV}^2$, which is not so easy to reconcile with our first
intuition from the naive quark model, namely, $\Delta \Sigma = 1$.

%%%%%%%%%%%%%%%%%%%%%%%%%%%%%%%%%%%%%%%%%%%%%%%%%%%%%%%%%%%%%%%%%%%%%%%%%%%%%%
\section{Summary}
\label{scope}

Our knowledge of the unpolarized and polarized structure functions has
improved considerably over the last twenty years or so and consequently
also, our basic understanding of the proton structure. The enormous
quantity of high statistics experimental data which is now available,
allows us to make rather precise tests of the QCD parton model and this
theoretical picture is nowdays established on very serious basis. Deep
inelastic phenomena is only one of the several areas of high energy
particle physics which are the testing grounds of perturbative QCD and
obviously, it has to be supplemented at least by electron-positron
annihilation and hard hadron-hadron collisions\cite{STE}.

For the unpolarized parton distributions, the flavor symmetry breaking
of the sea quarks ({\it i.e.\/} $d_{sea} > u_{sea}$) was only
discovered recently with the violation of the GSR and it should be
investigated more precisely for example in Drell-Yan lepton pair
production or in $W^\pm$ production\cite{BOU}. Our knowledge about the
gluon distribution, {\it i.e.\/} its small $x$ behavior and $Q^2$
dependence, is improving but a real experimental effort remains to be
done to put it on equal footing with the quark distributions. It
requires, for example, better measurements of $F_2^p (x, Q^2)$ in the
very small $x$ region and this makes HERA-DESY really exciting for the
coming years.

Concerning the polarized parton distributions significant progress have
been made and we have now the following results. The violation of the
EJSR is confirmed for the proton and for the neutron. The BSR is verified 
up to 10\% and
this important test must be strengthened at a higher level of accuracy.
The kinematic domain presently accessible is still too limited to
answer some important questions, in particular about the small $x$
behavior of $g_1$, the $Q^2$ evolution of the quark distributions and
about the, so far, elusive $\Delta G (x, Q^2)$. The perspective of having
polarized protons at HERA-DESY, which is seriously envisaged\cite{ZEU}, is
opening a new experimental window for answering these questions.
Finally let us mention the very exciting spin physics programme which
has been approved at RHIC-BNL and will start operating by 2001 as a polarized 
$pp$ collider\cite{SAI}. Hopefully
it will also shade some light on parton helicity distributions as well
as quark transversity distributions. A facility called eRHIC, which combines a
few GeV polarized electron beam onto one of the high energy polarized proton 
beam, is also under serious studies at BNL.

%%%%%%%%%%%%%%%%%%%%%%%%%%%%%%%%%%%%%%%%%%%%%%%%%%%%%%%%%%%%%%%%%%%%%%%%%%%%%%
%\begin{thebibliography}{999}

\end{document}